\documentclass[twocolumn]{aastex631}

\usepackage{amsmath}

\begin{document}

\title{The Mass of the Vela Pulsar Progenitor and the Age of the Vela-Puppis Complex}

\correspondingauthor{Jeremiah W.~Murphy}
\email{jwmurphy@fsu.edu}

\author[0000-0003-1599-5656]{Jeremiah W.~Murphy }
\affiliation{Department of Physics\\
Florida State University\\
77 Chieftan Way\\
Tallahassee, 32306, FL, USA}

\author{Andr\'es F. Barrientos}
\affiliation{Department of Statistics\\ 
Florida State University\\
117 N. Woodward Ave. \\
Tallahassee, 32306, FL, USA}

\author[0000-0001-8006-6365]{Ren\'e Andrae}
\affiliation{Max-Planck-Institut f{\"u}r Astronomie \\
K{\"o}nigstuhl 17 \\
Heidelberg, D-69117, Germany}

\author[0000-0001-8878-4994]{Joseph Guzman}
\affiliation{Department of Physics\\
Florida State University\\
77 Chieftan Way\\
Tallahassee, 32306, FL, USA}

\author[0000-0002-7502-0597]{Benjamin F.~Williams}
\affiliation{Department of Astronomy \\
University of Washington \\
Box 351580\\
Seattle, WA, 98195, USA}

\author[0000-0002-1264-2006]{Julianne J.~Dalcanton}
\affiliation{Center for Computational Astrophysics\\
Flatiron Institute\\
162 Fifth Ave
New York, NY, 10010, USA}
\affiliation{Department of Astronomy \\
University of Washington \\
Box 351580\\
Seattle, WA, 98195, USA}

\author[0000-0001-5530-2872]{Brad Koplitz}
\affiliation{
School of Earth \& Space Exploration\\
Arizona State University\\
781 Terrace Mall\\
Tempe, AZ, 85287, USA}



\begin{abstract}

The association of the Vela Pulsar with the Vela Supernova Remnant has long supported the hypothesis that core-collapse supernovae yield neutron stars, but its surrounding stellar population now offers new insights into progenitor evolution. By age-dating stars within 150 pc of the Vela Pulsar, we infer properties of its progenitor. 
These stars belong to the Vela-Puppis complex, revealing the region’s star formation history. While stellar population models with standard assumptions suggest a likely progenitor age and mass, these predictions are internally inconsistent with the observed population, indicating that something is missing in the standard modeling approach.
With those assumptions, there is very weak support for a $\lesssim$10 Myr old population, moderate support for a 40 Myr old population, and strong support for an intermediate age population around 65-100 Myrs old.  
The $\lesssim$10 Myr signal hinges on two peculiar O stars, which are unlike any others in the Vela-Puppis complex and imply nearly three times more main sequence stars than are observed.
The 40 Myr-old population is supported by 6 red supergiants (RSGs) and several Be stars; 
but this population is again marginally inconsistent with the observed distribution of main sequence stars.
The red giant (RG) and MS distributions are consistent with a 65-100 Myr old population.
We discuss several possible resolutions, emphasizing how binary evolution and/or very rapid rotation could resolve these discrepancies. Gaia parallaxes and {\it Stellar Ages} enable these results; {\it Stellar Ages} is a novel stellar population modeling algorithm that combines individual and population-level age inferences.
\end{abstract}



\section{Introduction} \label{sec:intro}

The association of the Vela Pulsar with its supernova remnant (SNR) was one of the first clear indications that core-collapse supernovae (CCSNe) produce neutron stars \citep{large1968}, and it remains a key system for studying supernova progenitors. As one of the nearest ($287^{+19}_{-17}$ pc) and best-studied pulsars \cite{dodson2003}, Vela provides a rare opportunity to infer the properties of its progenitor through the analysis of its surrounding stellar population. Traditional approaches assume that the progenitor evolved as a single massive star \citep{kochanek2022}, leading to constraints on its initial mass based on the age of coeval stars in its vicinity. However, recent advances in stellar population age-dating techniques \citep{guzman2025} challenge these assumptions, revealing inconsistencies that suggest a need to reconsider the roles of binary evolution and/or stellar rotation.

Using Gaia parallaxes and a refined statistical framework, we find that the distribution of massive stars within 150 pc of the Vela Pulsar does not conform to expectations based on standard modeling assumptions (single-star, relatively low rotation, standard IMF, etc.).  While previous analyses suggested a progenitor mass in the range of 8.1–10.3 M$_{\odot}$ \citep{kochanek2022}, our results indicate a more complex stellar history, with the most likely scenarios involving binary evolution or very rapid rotation. This finding aligns with growing evidence that binary interactions play a significant role in shaping the final stages of massive star evolution and may be critical in determining which stars explode as supernovae \citep{smith2011,sana2012,demink2014,zapartas2017,eldridge2017,renzo2019,offner2023}.

Furthermore, rotation may be an essential factor in reconciling the observed stellar population with theoretical predictions. The presence of peculiar O stars and Be stars suggests a history of rapid rotation or binary interactions, complicating the traditional mass-age relationship used to infer progenitor properties. Our results underscore the need to revisit standard assumptions in stellar population modeling when interpreting supernova progenitors and their environments.  Additionally, these findings have significant implications for understanding the history and structure of the Vela-Puppis OB association complex, shedding light on its star formation history and the distribution of massive stars within it.

A key advance in this study is the use of the {\it Stellar Ages} statistical technique, which uniquely combines the specificity of inferring ages for individual stars with the ability to provide population-level constraints. Historical approaches have typically inferred properties either for individual stars or for entire populations separately. By integrating both perspectives, {\it Stellar Ages} enables the detection of internal inconsistencies that were previously obscured, revealing crucial insights into the evolutionary history of the Vela Pulsar progenitor.

This paper presents a revised age-dating analysis of the stellar population surrounding the Vela Pulsar, demonstrating that its progenitor likely experienced significant binary evolution and/or very rapid rotation. In Section~\ref{sec:data}, we describe the selection and verification of the stellar population. Section~\ref{sec:model} details our Bayesian inference framework for age-dating. Section~\ref{sec:results} presents the results, highlighting that the data are inconsistent with the standard assumptions in modeling stellar populations. Section~\ref{sec:resolutions} explores several ways to resolve these inconsistencies and highlights how binary interactions and/or very rapid rotation could resolve these inconsistencies.  Section~\ref{sec:progenitor} discusses the broader impact of these findings on CCSN progenitor studies, and finally, we summarize our conclusions in Section~\ref{sec:conclusions}.

\section{Data Selection, Correction, and Verification}
\label{sec:data}

The exquisite astrometric precision of VLBI and Gaia makes it possible to age-date the stellar population near the Vela Pulsar with unprecedented accuracy.  The Vela Pulsar itself has been well studied for decades, with its position and motion precisely determined by \citet{dodson2003} using VLBI astrometry.  Their measurements established its sky coordinates at
($\alpha = 08^{\rm h}35^{\rm m}20.61149^{\rm s} \pm 0.00002^{\rm s}$ and $\delta = -45^{\circ}10^{\prime}34.8751^{\prime \prime} \pm 0.0003$ in the 2000.0 epoch),  
corresponding to galactic coordinates of $(l,b) = (263.55183143^{\circ}, -2.78731237^{\circ})$. 
Its parallax of $3.5 \pm 0.2$ mas places it at a distance of
287$^{+19}_{-17}$ pc,
positioning it $\sim$14 pc below the galactic plane, well within the Milky Way's thin disk.

Unlike many pulsars, which are kicked to high velocities at birth and quickly ejected from their natal environments  \citep{hansen97}, the Vela Pulsar moves at a modest transverse velocity of $61 \pm 2$ km s$^{-1}$ relative to the local standard of rest \citep{dodson2003}.  This is a crucial point: its low velocity increases the likelihood that it remains embedded within its birth association, meaning that the nearby stellar population may provide a direct window into the conditions of its formation.  Using its proper motion and its current offset of $25 \pm 5$ arcmin from the center of the Vela SNR \citep{aschenbach1995}, \citep{kochanek2022} estimated its age to be $\sim$20,000 years. This implies that, since its explosion, the pulsar has traveled only $\sim$1.5 pc—further strengthening the case that it remains within its birth environment. Taken together, these factors make the Vela Pulsar a particularly valuable test case for studying core-collapse supernovae (CCSNe) in the context of their progenitor populations.

To identify stars that may have formed alongside the Vela Pulsar’s progenitor, we turn to the unparalleled astrometric precision of Gaia Data Release 3 \citep[DR3]{gaiamission2016,gaiadr32022}. 
Our goal is to select stars that are both spatially and kinematically consistent with having originated in the same star-forming event.
Similar to \citet{kochanek2022}, we define our search region as a 150 pc radius sphere centered on the pulsar’s present location. Given that the pulsar itself has moved less than 2 pc since its explosion, this region is large enough to encompass the bulk of its natal stellar population while minimizing contamination from unrelated field stars.

Our selection criteria focus on three primary constraints. First, we require sky coordinates within 27$^\circ$ of the pulsar, ensuring coverage of known OB associations in the broader Vela-Puppis complex. 
Second, we select stars with parallaxes between 2.27 and 7.14 mas, corresponding to distances of 140–440 pc—slightly wider than the pulsar’s own parallax uncertainties to account for depth effects within the association.
Third, we impose an apparent BP-band magnitude limit of $\le$10 to prioritize stars bright enough for robust age determination while reducing contamination from faint, background stars with high relative parallax uncertainties.
To further refine our sample, we convert Gaia apparent magnitudes to absolute magnitudes, correcting for parallax, zero-point systematics \citep{lindegren2021}, and extinction.  For example, the absolute magnitude in the red band (RP) is
\begin{equation}
M_{RP} = G_{RP} - 5 log_{10}(\frac{1000}{10 (\varpi - z_{\rm pt})}) - A_{RP} \,  
\end{equation}
where $G_{RP}$ is the Gaia apparent magnitude, $\varpi$ is the parallax in milliarcseconds, $z_{\rm pt}$ is the zero point, and $A_{RP}$ is the GSP-Phot extinction.  
We then impose a final selection criterion: stars must have absolute magnitudes in both BP and RP brighter than 0, roughly corresponding to a lower mass threshold of $\sim$3 M$_{\odot}$ on the main sequence (MS). This ensures that our sample is dominated by the high-mass stars most relevant for CCSN progenitor studies.

Although extinction is relatively low in this region, its effects on magnitudes must still be accounted for. There are two common approaches to handling extinction when modeling color-magnitude diagrams (CMDs). One approach is to apply extinction corrections to each individual star based on known or measured values. However, when extinction is unknown for individual stars, it is necessary to infer an extinction distribution for the stellar population as a whole. {\it Stellar Ages} supports both approaches: it can either infer an extinction distribution or apply corrections based on prior extinction information \citep{guzman2025}. In previous age-dating of the Vela Pulsar’s stellar population, \citet{kochanek2022} used the extinction distribution approach. In this study, we instead use Gaia extinction estimates to correct each individual star. These estimates are derived from the General Stellar Parameterizer for Photometry (GSP-Phot) models, which use Gaia BP and RP spectra to infer a range of stellar parameters, including extinction \citep{andrae2023}.  For the vast majority of stars in our sample, extinction is small: over 90\% have $A_{\rm RP}$ and $A_{\rm BP} < 0.2$, with a median extinction below 0.1. These low values indicate that extinction corrections are unlikely to introduce significant systematic uncertainties in our analysis.

One challenge in assembling a complete sample of massive stars is that Gaia DR3 has known incompleteness at the bright end, particularly for stars with 
$G < 3$ (see Gaia DR3 documentation, Sec. 14.3.1 and Fig. 14.38).
To compensate for this, we supplement our sample with bright stars from the Hipparcos catalog that are missing from Gaia DR3. Using the same spatial and parallax constraints as before, we select Hipparcos stars within 150 pc, then transform their Johnson-Cousins V and I magnitudes into Gaia BP and RP magnitudes using the color-magnitude transformations of \citet{riello2021}. To avoid duplication, we cross-match with Gaia using the Hipparcos-Gaia catalog of \citet{brandt2021}, removing any stars that already have a Gaia DR3 counterpart within 3$^{\prime \prime}$. This process adds 13 Hipparcos-only stars to our sample, six of which are bright evolved stars—ensuring that our sample remains representative of the full high-mass stellar population in this region.

\begin{figure}
\includegraphics[width=\columnwidth]{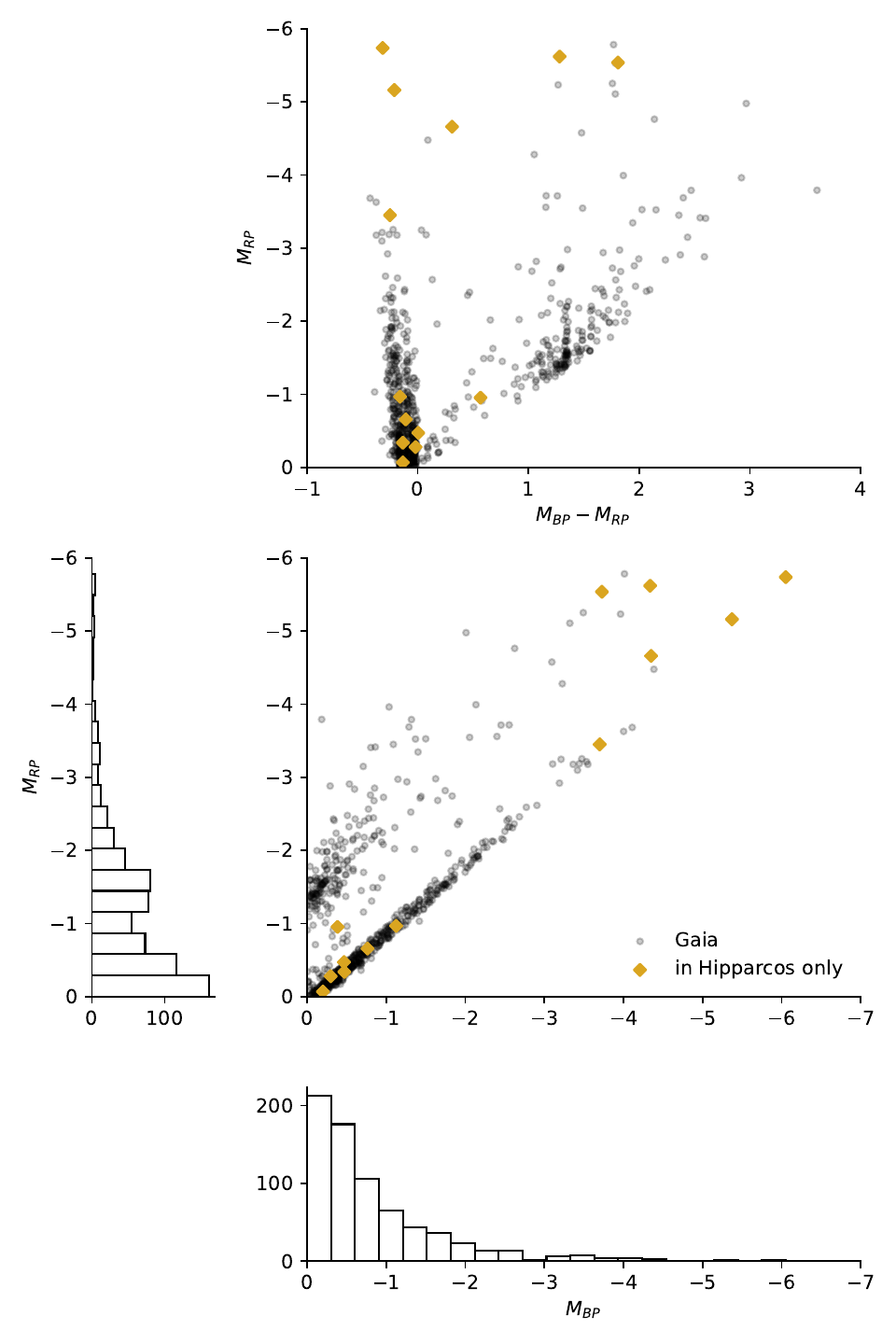}
\caption{The absolute magnitudes of all stars within 150 pc of the Vela Pulsar. The top-right panel displays the color-magnitude diagram (CMD), showing the Gaia red-band absolute magnitude ($M_{RP}$) versus the Gaia color index ($M_{BP} - M_{RP}$). Instead of modeling the CMD directly, {\it Stellar Ages} models individual absolute magnitudes to avoid introducing artificial correlations between the abscissa and ordinate. The middle-right panel presents the absolute magnitude-magnitude diagram, illustrating the distribution of stars in $M_{RP}$ vs. $M_{BP}$. Both panels include all stars with apparent magnitudes $\leq 0$ in both filters. While Gaia DR3 provides the most extensive collection of astrometric solutions at these distances, it lacks reliable solutions for the brightest stars. To address this, we supplement the sample with stars from the Hipparcos catalog (gold points), ensuring completeness at the bright end. The final dataset consists of 705 stars from Gaia DR3 and 13 additional stars from Hipparcos.}\label{fig:gaiahip_select}
\end{figure}

Figure~\ref{fig:gaiahip_select} presents the absolute magnitudes of all selected stars, distinguishing between those from Gaia DR3 (black points) and those added from Hipparcos (gold points). The top panel displays the color-magnitude diagram (CMD) using $M_{\rm RP}$~vs.~$M_{\rm BP}-M_{\rm RP}$, while the middle-right panel plots absolute magnitudes in RP against BP. The lower-left and lower-right panels provide histograms of the absolute magnitudes in each band. These visualizations serve two purposes: first, they demonstrate that our sample spans a broad and representative range of stellar types, and second, they highlight the importance of including the Hipparcos stars to ensure completeness at the bright end.

In addition to the CMD, we also present the absolute magnitude-magnitude diagram (middle-right panel), where $M_{\rm RP}$ is plotted directly against $M_{\rm BP}$.  This alternative representation is particularly useful for our analysis because the axes in a CMD are inherently correlated. By modeling absolute magnitudes independently rather than using the CMD directly, {\it Stellar Ages} avoids introducing artificial correlations into the inference process. This approach ensures that age estimates are driven by the intrinsic properties of the stars rather than statistical artifacts introduced by correlated uncertainties.  

Several key features of the stellar population are evident in the magnitude-magnitude diagram. The lower band consists of main-sequence (MS) stars, while the upper band is composed of red giants (RGs) and red supergiants (RSGs). Most of the RGs are old, low-metallicity halo stars, distinct from the younger massive stars of interest. At the bright end, we see a small population of luminous giants and Ib supergiants, but notably, there are no bright Ia supergiants (see Section~\ref{sec:simbad}). This absence may be significant when interpreting the evolved stellar population in relation to the progenitor of the Vela Pulsar.

In total, our final sample consists of 718 stars: 705 from Gaia DR3 and 13 from Hipparcos. Given the large size of the Gaia sample, the addition of 13 Hipparcos stars—while valuable for capturing the full range of bright evolved stars—does not significantly alter the statistical inference for the overall stellar population. However, their inclusion is important for ensuring that we account for the most luminous members of the Vela-Puppis complex, which are critical for constraining the age distribution.

Uncertainty in absolute magnitudes is dominated by parallax errors. However, because the stars in our sample are relatively nearby, parallax uncertainties remain small. More than 99\% of the stars have relative parallax uncertainties ($\sigma_{\varpi}/\varpi$) below 0.05, with most below 0.01.  Even in the most extreme cases, uncertainties remain below 0.2. This high precision ensures that our inferred absolute magnitudes—and by extension, our age-dating analysis—are not significantly impacted by parallax errors.

\section{Age-dating the Stellar Population using Single-Star Models}
\label{sec:model}

To determine the age distribution of the stellar population surrounding the Vela Pulsar, we use {\it Stellar Ages} \citep{guzman2025}, a statistical framework that infers the distribution of ages, metallicities, rotation rates, and extinction properties based on individual stellar magnitudes. This approach enables us to extract age constraints not just for the population as a whole, but for individual stars as well. Unlike traditional methods that rely on binned color-magnitude diagrams (CMDs), {\it Stellar Ages} models the joint probability distribution of absolute magnitudes, avoiding information loss from binning and minimizing biases due to correlated uncertainties. By leveraging this method, we can test whether the observed stellar population is consistent with standard single-star evolution or if additional effects, such as binary evolution or rapid rotation, are required to explain the data.

Most traditional age-dating techniques infer the star formation history by counting the number of stars in discrete CMD bins \citep{dolphin2002,dolphin2012,dolphin2013}. These methods assume that the observed number of stars in each bin follows a Poisson distribution, with the likelihood function constructed as the joint probability of all bins. While this approach is effective for large stellar populations, such as entire galaxies, it is suboptimal for the analysis of relatively small samples of bright, evolved stars. Since stars spend $\sim$90\% of their lifetimes on the main sequence, most stars in a given population are main-sequence stars, leading to small-number statistics for evolved stars. This limits the ability of binned CMD methods to effectively constrain the ages of the most massive stars—precisely the ones most relevant for supernova progenitor studies.


Rather than modeling the number of stars in CMD bins, {\it Stellar Ages} directly infers the ages of individual stars by leveraging the known probability density functions of stellar magnitudes. This approach fundamentally shifts the statistical question: instead of asking, "How many stars do we expect at a given age?", we ask, "What is the most likely age of each observed star?" This distinction is critical because it maximizes the information contained in the brightest stars. For example, a star with an absolute magnitude of 
$M_{\rm F814W} = -8.6$ must be younger than 10 Myr, implying a minimum initial mass of at least 19 M$_{\odot}$. Even a handful of such stars can provide a strong constraint on the population’s age, whereas traditional methods would struggle due to Poisson uncertainties in sparsely populated CMD bins. 
However, instead of simply treating each star’s age as fully independent, {\it Stellar Ages} constrains individual age inferences within the broader statistical framework of the population, ensuring self-consistency across the entire sample.
This allows for a more precise and robust determination of the stellar age distribution while mitigating biases that arise when considering stars in isolation.

In general, {\it Stellar Ages} infers the posterior distribution for age ($t$), metallicity ([M/H]), initial rotation as a fraction of the critical rotation ($v_{\rm ini}$), and the mean extinction, $\tilde{A}_{V}$.  Because GSP-Phot already provides extinction corrections, we do not need to infer the mean extinction. Therefore, the model parameters are $\theta = \{t, \rm{[M/H]}, v_{\rm{ini}} \}$.  Schematically, the posterior distribution is:
\begin{equation}
P(\theta | D) \propto \mathcal{L}(D | \theta) P(t) P(\rm{[M/H]}) P(v_{\rm{ini}}) \, .
\end{equation}
The data ($D$) are the set of magnitudes for all stars.  The modeled magnitude in band ($a$) of each star is
\begin{equation}
    m_{a} = \tilde{m}_{a}(M,t,\rm{[M/H]},v_{\rm{ini}}) + A_a + e_{a} \, ,
    \label{eq:maga}
\end{equation}
where $\tilde{m}_a$ is the magnitude predicted by stellar evolution models as a function of $t$, [M/H], and the initial mass ($M$) of the star \citep[PARSEC v1.2]{bressan2012,chen2014,chen2015,fu2018}.  $e_a$ is a random error, drawn from a Gaussian distribution with width $\sigma_a$; $\sigma_a$ is the quadrature sum of the magnitude and parallax uncertainties.  $A_{a}$ represents the extinction parameter in band $a$; again GSP-Phot provides the extinction.  In this analysis, we model two magnitudes, which are the Gaia absolute magnitudes BP and RP.


The joint likelihood of observing a star with magnitudes $m_a$ and $m_b$ is 
\begin{multline}
 \mathcal{L}(m_a,m_b|\theta) = \int p(m_{a}|\tilde{m}_{a})p(m_{b}|\tilde{m}_{b})p(\tilde{m}_{a}|\theta,M) \\
 p(\tilde{m}_{b}|\theta,M) p(M) dM d\tilde{m}_{a} d\tilde{m}_{b} \, . 
\end{multline}
This likelihood is the joint probability density function for a given age and metallicity.  $p(m_{a}|\tilde{m}_{a})$ and $p(m_{b}|\tilde{m}_{b})$ are Gaussian distributions whose widths represent the observational uncertainties.  $p(\tilde{m}_a | \theta, M)$ and $p(\tilde{m}_b | \theta, M)$ are delta functions and come directly from stellar evolution.  Since $\tilde{m}_a$ and $\tilde{m}_b$ are not analytic, there is no closed analytic form for the joint probability density function.  To approximate the likelihood, we note that the likelihood is the expectation of $p(m_a | \tilde{m}_a)$ and $p(m_b | \tilde{m}_b)$ with respect to the initial mass $M$.  Therefore,
\begin{multline}
    \mathcal{L}(m_a,m_b|\theta) = 
    E_M \left [ p(m_a | \tilde{m}_a(\theta, M)) 
    p(m_b | \tilde{m}_b(\theta, M)) \right ] \\
    \approx \frac{1}{N_M} \sum^{N_M}_\ell 
    p(m_A | \tilde{m}_A(\theta, M^{(\ell)}))
    p(m_B | \tilde{m}_B(\theta, M^{(\ell)})) \, ,
\end{multline}
where each $M^{(\ell)}$ is a draw from $P(M)$, the initial mass function.  To minimize sampling noise, we use perfect sampling instead of random sampling when drawing from this distribution.


Real stellar populations will likely be a mixture of populations with different ages and metallicities. Therefore, we propose a mixture data-generating model, or a weighted sum of the individual joint probability density functions:
\begin{multline}
\label{mix_model}
    \mathcal{L}(m_A,m_B|\{w_{t,z,v}\}^{T,Z,V}_{t=1,z=1,v=1}) \\
    = \sum^T_t \sum^Z_z \sum^V_v w_{t,z,v} \mathcal{L}(m_A,m_B|\theta_{t,z,v}) \, ,
\end{multline}
where $w_{t,z,v}$ are weights for age index $t$, metallicity index $z$, and initial rotation index $v$.


To infer the weights, we employ Bayesian statistics. Since sampling from the posterior distribution is not straightforward, we employ a Gibbs sampler algorithm that relies on the introduction of latent variables called labels ($r_i$). These labels assign specific age,  metallicity, and rotation values to each star. Thus, the mixture model (\ref{mix_model}) is equivalently
\begin{eqnarray}\nonumber
(m_{A,i},m_{B,i} | r_i = (t', z',v'))  &\sim& \mathcal{L}(m_A,m_B|\theta_{t',z',v'}) \\ \label{mix_model_ri}
r_i = (t', z',v') | \{w_{t,z,v}\}^{T,Z,V}_{t=1,z=1,v=1} & \sim & w_{t',z',v'} \quad (i = 1...N_{\rm stars}) \, .
\end{eqnarray}
We complete the model specification by assigning a prior distribution to the weights. Specifically, we assume that  
\begin{equation}\label{prior_weights}
    \{ w_{t,z,v} \}^{T,Z,V}_{t=1,z=1,v=1} \sim {\rm Dirichlet}(1, ...,1) \, 
\end{equation}
where $\rm{Dirichlet}(1, ...,1)$ denotes a $(T \times Z \times V)$-dimensional Dirichlet distribution. Under model (\ref{mix_model_ri}) with prior (\ref{prior_weights}), the Gibbs sampler iterates between updating the labels and weights.

To update the labels, we sample from the conditional distribution 
\begin{multline}
P(r_i = (t',z',v')| \{ w_{t,z,v} \}^{T,Z,V}_{t=1,z=1,v=1}) \\ = \frac{w_{t',z',v'} p(m_{a,i},m_{b,i} | \theta_{t',z',v'}) }{ \sum_{t} \sum_{z} \sum_v w_{t,z,v} p(m_{a,i},m_{b,i}|\theta_{t,z,v})} \,  \quad (i = 1...N_{\rm stars}) \, .
\end{multline}
To update the weights, we first compute the number of $r_i$, $i = 1...N_{\rm stars}$, that are equal to $(t',z',v')$ denoting the resulting count as $N(t=t',z=z',v=v')$. Then, we update the weights by sampling from the Dirichlet distribution:
\begin{multline}
    \{ w_{t,z,v} \}^{T,Z,V}_{t=1,z=1,v=1} | \{r_i\}_{i=1}^{N_{\rm stars}} \\
    \sim {\rm Dirichlet}[1+N(t=1,z=1,v=1) \\,...,1+N(t=T,z=Z,v=V)] \, .
\end{multline}

 For more discussion on the techniques and validation tests of {\it Stellar Ages}, see \citet{guzman2025}.

\section{The Stellar Population is Inconsistent with Single-Star Evolution}
\label{sec:results}

\begin{figure}
\includegraphics[width=\columnwidth]{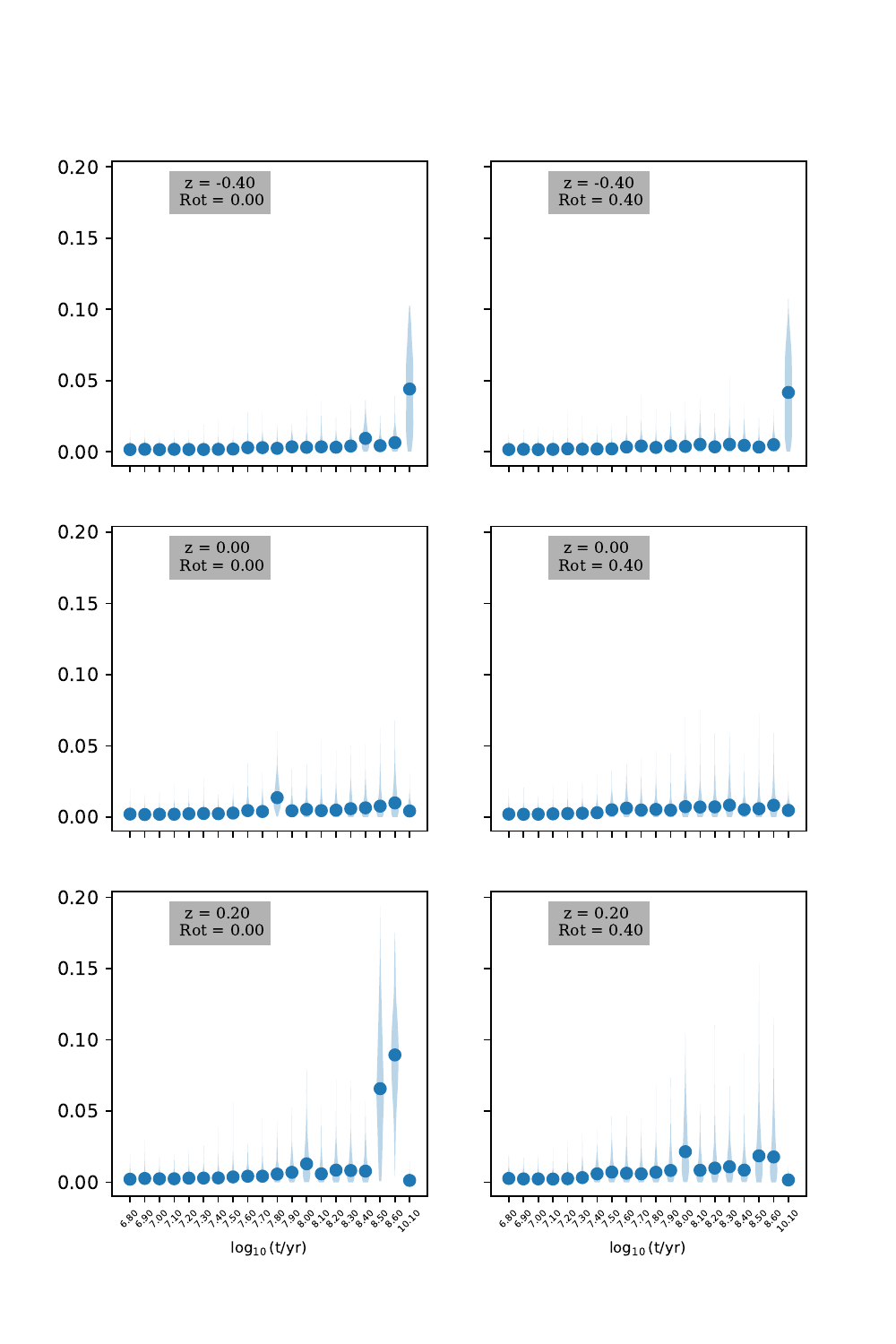}
\caption{
Inferred ages, metallicities, and rotation weights for the stellar population within 150 pc of the Vela Pulsar, based on MCMC sampling. The stellar isochrones used in this inference are from the MIST single-star evolutionary models. This violin plot summarizes the posterior distribution of weights, where the central dot represents the median weight, and the width of the violin reflects the distribution of sampled values.  While the brightest blue stars superficially appear consistent with a young population, the MCMC inference instead reveals three distinct age components: (1) a dominant population at $\log_{10}(t/{\rm yr}) \ge 7.8$ ($\sim$63 Myr or older), (2) a weaker peak around 40 Myr, and (3) a very weak component at $\le$10 Myr, suggesting that the youngest population is sparse and unlikely to represent a true coeval group.}\label{fig:weights_grid}
\end{figure}

Figure~\ref{fig:weights_grid} presents the posterior distribution for the age-metallicity-rotation weights, $w_{t,z,v}$, for the stellar population within 150 pc of the Velar Pulsar.  Each panel shows the inferred probability density as a function of log age 
$\log_{10}(t/\rm{yr})$, with separate panels for different metallicity [M/H] and rotation ($v_{\rm crit}$).
The Bayesian inference reveals three distinct age components: (1) a dominant population at $\log_{10}(t/{\rm yr}) \ge 7.8$ ($\sim$63 Myr or older, (2) a weaker peak around 40 Myr, and (3) a very weak component at $\lesssim$10 Myr. While the 40 Myr population retains some statistical weight, the youngest population is sparsely populated, raising doubts about whether it represents a genuine coeval group.

To analyze the stellar population age distribution, we run an MCMC with 1000 steps. The solution converges within $\sim$20 steps, so we discard the first 50 steps as burn-in. Since MCMC draws are correlated from one step to the next, we select only every 10th step to ensure statistical independence, resulting in 95 unique MCMC draws.  Figure~\ref{fig:weights_grid} shows the posterior distribution of age weights, $w_{t,z,v}$, where the middle dot represents the median weight, and the width of the violin is proportional to the distribution of samples.

\begin{figure}
\centering
\includegraphics[width=\columnwidth]{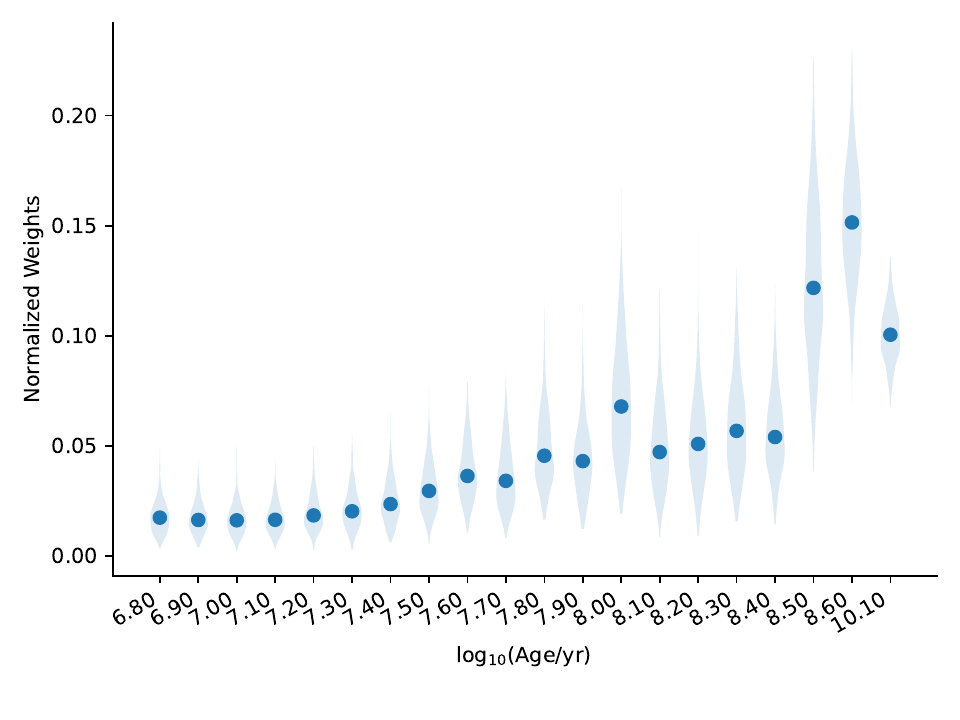}
\caption{
The posterior distribution of age weights, marginalized over metallicity ($[M/H]$) and initial rotation. While the brightest stars superficially suggest young ages ($\le$10 Myr and $\sim$40 Myr), the posterior distribution assigns very little weight to the youngest ages and only moderate weight to the $\log_{10}(t/{\rm yr})=7.6$ (40 Myr) component. This indicates that only a small fraction of stars support these young ages. Instead, the dominant population is at $\log_{10}(t/{\rm yr}) \ge 7.8$ ($\ge$ 63 Myr). Further examination of the inference reveals that the $\log_{10}(t/{\rm yr})=7.6$ solution is not only weak but also inconsistent with the data (see Figures~\ref{fig:ageperstar}-\ref{fig:threeages}).
}\label{fig:weights}
\end{figure}

Figure~\ref{fig:weights_grid} provides an initial hint of the discrepancy, but Figure~\ref{fig:weights} makes it clearer by marginalizing over metallicity and rotation. The youngest ages ($\le$10 Myr) contribute almost no weight to the posterior distribution, while the strongest support remains at 63–100 Myr, with a secondary peak at 40 Myr. These results indicate that if a young ($\le$10 Myr) population exists, it must be extremely sparse—far too sparse to account for the presence of the two bright O stars, which, under standard single-star evolution, would be expected to be accompanied by a much larger number of lower-mass main-sequence (MS) stars.

\begin{figure}
\includegraphics[width=\columnwidth]{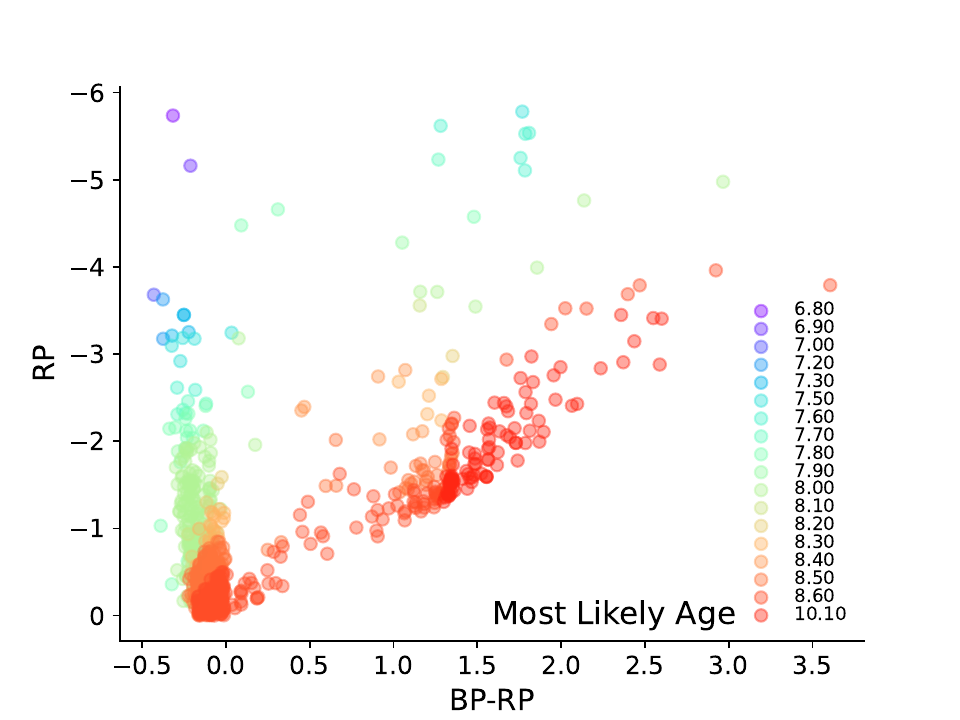}
\caption{The color and absolute magnitude diagram and the most likely inferred age for each star. In addition to inferring age weights for the entire population, \textit{Stellar Ages} estimates an individual most likely age for each star, enabling a direct examination of which stars contribute to which age groups. This approach reveals a key discrepancy: while several bright, evolved stars have inferred ages of $\log_{10}(t/{\rm yr}) \leq 7.7$ ($\le$ 50 Myr), there are almost no main-sequence (MS) stars with similar inferred ages.  Figures~\ref{fig:likelihood_data}~and~\ref{fig:threeages} further quantify and illustrate this inconsistency, demonstrating that the inferred age distribution of bright evolved stars is incompatible with the MS population.}\label{fig:ageperstar}
\end{figure}

Figure~\ref{fig:ageperstar} adds further context by color-coding each star on the CMD with its most likely age. While several bright, evolved stars appear consistent with young ages ($\log_{10} (t/{\rm yr}) \le 7.7$), there are strikingly few MS stars with similar ages. This discrepancy suggests that the young-looking evolved stars are not coeval with a young MS population. Instead, they may have undergone alternative evolutionary pathways, such as binary evolution or extreme rotation, making them appear younger than they actually are.

\begin{figure}
\includegraphics[width=\columnwidth]{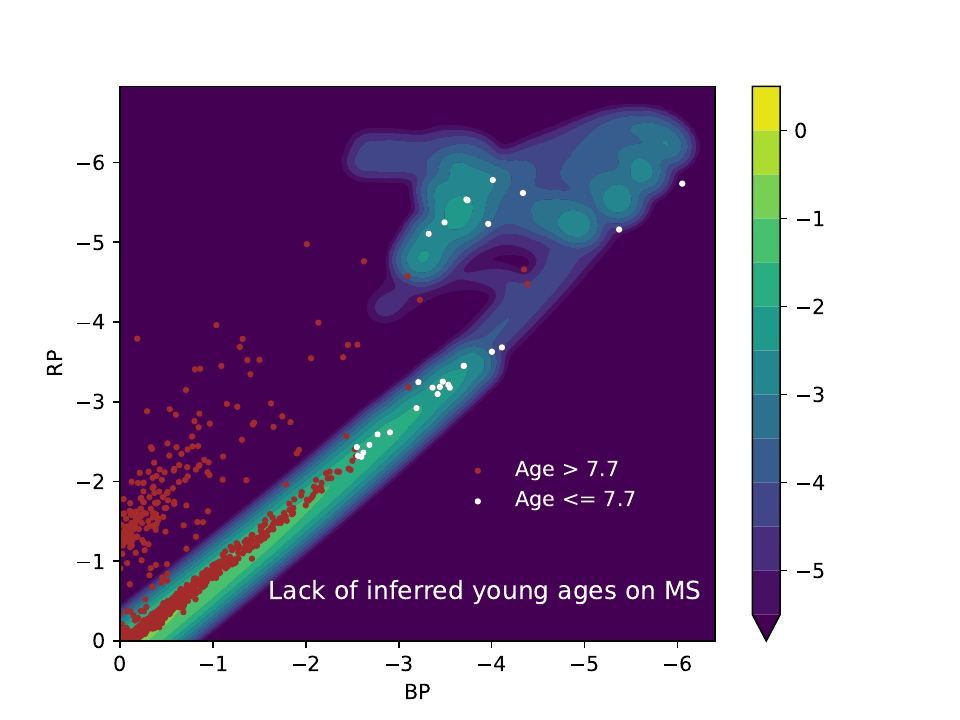}
\caption{Comparison between the observed data (absolute magnitudes) and the weighted model. The background color map represents the sum of all model likelihoods, weighted by the inferred age distribution from Figure~\ref{fig:weights}. The brown dots represent stars with most likely ages $\log_{10}(t/{\rm yr}) > 7.7$ (older population), while the white dots represent stars with most likely ages $\log_{10}(t/{\rm yr}) \leq 7.7$ (younger population). Although the brightest evolved stars appear young, the best-fitting model assigns almost no MS stars to these young ages. This is a key indication that the stellar population is inconsistent with single-star evolution alone.}\label{fig:likelihood_data}
\end{figure}

\begin{figure*}
\includegraphics[width=\textwidth]{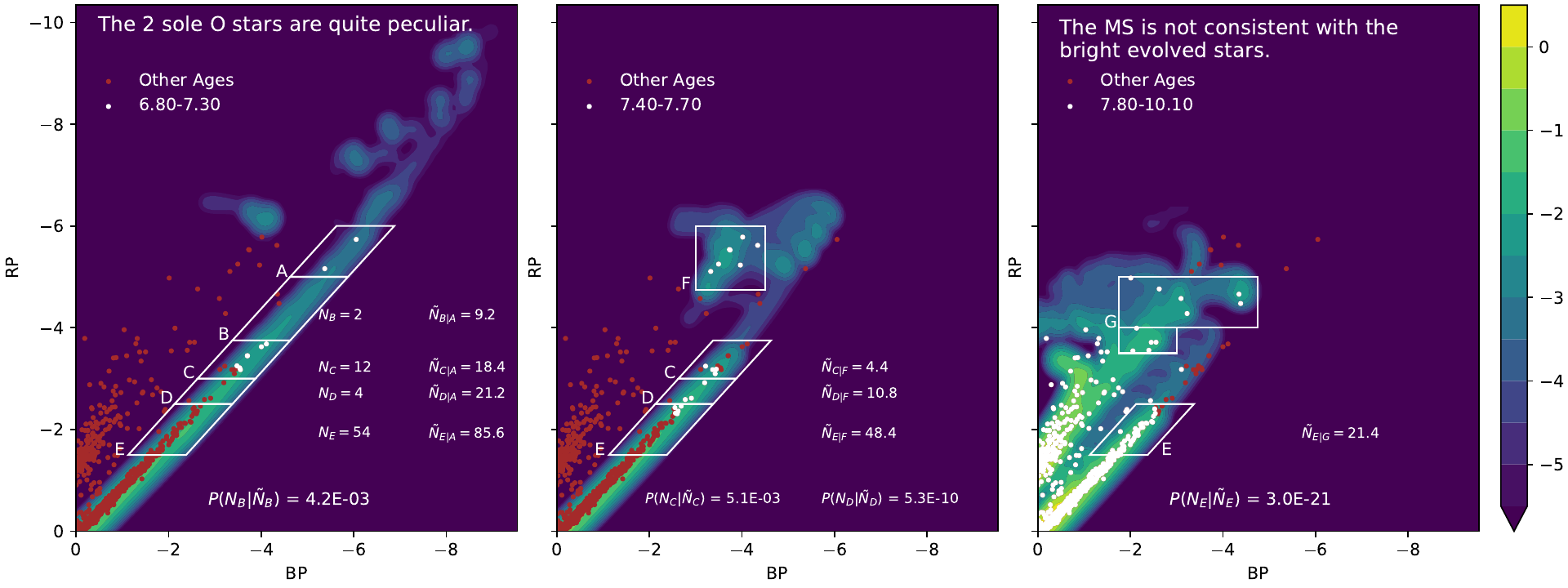}
\caption{These panels illustrate the inconsistency between the models and the observed ratio of bright evolved stars to MS stars. The white dots represent stars with inferred ages of $\log_{10}(t/{\rm yr}) = 6.80 - 7.3$ (left), $7.4 - 7.6$ (middle), and $\geq 7.80$ (right), while the brown dots represent stars of all other ages. The background color map shows the age-weighted model likelihoods. The figure is divided into labeled regions (A–G), where Region A contains the two peculiar O stars, Region F contains the six RSGs, and Region G contains the 11 bright RGs. The expected number of MS stars in each region is computed from the weighted model; for example, in Region B, $\tilde{N}_{B|A} = 9.2$ stars are expected based on the two O stars, but only $N_B = 2$ are observed. The probability values at the bottom quantify the discrepancy between the expected and observed number of stars. The left panel highlights the uniqueness of the two O stars, while the middle and right panels show that the bright evolved stars are inconsistent with standard single-star evolution under a Kroupa IMF.}\label{fig:threeages}
\end{figure*}

Figures~\ref{fig:likelihood_data}~and~\ref{fig:threeages} solidify the inconsistency by comparing the expected and observed number densities of stars across different magnitude regions. The key result is that the stellar population lacks the expected number of MS stars needed to support the inferred bright, evolved stars under a standard IMF.

Figure~\ref{fig:likelihood_data} compares the weighted model with the observed data to assess whether the young-looking evolved stars are supported by a corresponding young MS population. The background color map represents the weighted likelihood for the youngest ages, constructed by summing the likelihoods of all stellar models with ages 
$\log_{10}(t/{\rm yr}) \le 7.7$, weighted by the inferred age distribution from Figure~\ref{fig:weights}. The brown dots represent stars whose most likely age is older than 
$\log_{10}(t/{\rm yr})=7.7$, while the white dots represent stars with most likely ages younger than this threshold.

At first glance, the weighted model appears broadly consistent with the observed data, particularly in terms of the main-sequence turnoff location. However, a closer inspection reveals a fundamental inconsistency: the number densities of stars throughout the CMD do not match between the data and the model. While the brightest evolved stars exhibit colors and luminosities that suggest young ages, the best-fitting model assigns almost no MS stars to these ages. This discrepancy highlights a key issue: if the young-looking evolved stars truly originated from a coeval population, there should be a substantial number of corresponding MS stars at lower masses. The fact that these stars are largely absent in the best-fitting model suggests that the evolved stars do not belong to a single-age population, reinforcing the need for alternative evolutionary scenarios.

Figure~\ref{fig:threeages} most clearly illustrates the inconsistency between the expected and observed number densities of MS stars. This figure compares the actual number of stars in different regions of the magnitude-magnitude diagram to the number expected under a standard single-age stellar population model. The discrepancy is stark: the observed distribution of MS stars is fundamentally at odds with predictions from single-star evolution.

Each panel of Figure~\ref{fig:threeages} highlights a different age range. The left panel focuses on the youngest stars ($\log_{10}(t/{\rm yr})=6.8 - 7.3$), shown in white. The middle panel highlights stars with ages consistent with the red supergiants (RSGs), spanning $\log_{10}(t/{\rm yr})=7.4 - 7.7$. The right panel emphasizes the oldest stars ($\log_{10}(t/{\rm yr}) \ge 7.8$), which align with the brightest red giants (RGs).

To quantify the discrepancy, we divide the magnitude-magnitude diagram into several quadrilateral regions labeled A through G. Regions A through E span different parts of the main sequence (MS), covering a range of stellar luminosities. Region F contains the six RSGs, while Region G includes the eleven brightest RGs, which are best fit by ages $\log_{10}(t/{\rm yr}) = 7.8 - 8.0$.

The left panel of Figure~\ref{fig:threeages} compares the actual number of stars in each main-sequence region to the number expected if the two O stars in Region A were part of a coeval, young population. The observed star counts are denoted $N_B$, $N_C$, etc., while the expected counts—computed from the best-fitting model—are denoted $\tilde{N}{B|A}$, $\tilde{N}{C|A}$, and so forth. If the O stars were truly part of a young, single-age population, then Regions B through E should contain substantially more MS stars than are observed.

We evaluate this inconsistency using two complementary approaches. The first is a straightforward forward-likelihood calculation: we compute the probability of observing $N$ stars in each of Regions B–E, assuming the expected number $\tilde{N}$ derived from the best-fitting model. While intuitive and easy to compute, this approach treats the number of evolved stars in Regions A, F, and G as fixed inputs, without accounting for the statistical uncertainty that comes from their small sample sizes. To address this, we also perform a Bayesian posterior inference: we estimate the distribution of expected stars in Regions B–E, conditioned on the observed number of O stars, RSGs, and RGs. This second approach propagates the dominant source of uncertainty and provides a more conservative, but arguably more robust, assessment of model consistency.

Let us first consider the forward-likelihood approach.  A striking example of the discrepancy appears in Region B, which lies adjacent to the O stars in Region A.  Based on the standard single-age model and the presence of two O stars in Region A, we would expect $\tilde{N}_{B|A} = 9.2$ stars in region B.  However, only $N_B = 2$ are observed. Assuming a Poisson distribution, the probability of such a shortfall is $P(N_B | \tilde{N}_B) = 4.2 \times 10^{-3}$. If we exclude even these two stars—since they may not be MS stars—the probability of observing zero becomes $10^{-4}$. 
In either case, if the O stars are interpreted as members of a young coeval population, the lack of accompanying MS stars is highly improbable, suggesting instead that these O stars are outliers or evolved through nonstandard pathways.

The middle panel of Figure~\ref{fig:threeages} further emphasizes the inconsistency under the forward-likelihood approach by showing that Regions C and D also contain significantly fewer MS stars than expected. Although the O stars in Region A and the RSGs in Region F represent different age populations, both should contribute MS stars to these regions under standard single-star evolution and a standard IMF. In Region C, the combined expected number is $\tilde{N}_C = \tilde{N}_{C|A} + \tilde{N}_{C|F} 22.8$, yet only 12 stars are observed. The probability of this shortfall, assuming Poisson statistics, is $P(12|22.8) = 5 \times 10^{-3}$, reinforcing the statistical tension seen in Region B.

The discrepancy is even more pronounced in Region D, where the expected number of MS stars is $\tilde{N}_D = 32$, yet only 4 stars are observed, despite an expectation of . The corresponding Poisson probability is $\sim 10^{-10}$, underscoring the implausibility of such a shortfall under standard single-star evolution.  This trend continues in Region E, which should receive contributions from all three age groups—O stars, RSGs, and RGs -- resulting in an expected count of $\tilde{N}_E = 155.4$ stars.  However, only $N_E = 54$ stars are observed.
Taken together, these calculations point to a consistent conclusion: the presence of the two O stars is difficult to reconcile with the observed MS population, strongly suggesting that they are not part of a typical coeval population governed by standard stellar evolution.

Given this extreme inconsistency, one possibility is that the O stars are runaways—stars that originated in a different association and have since migrated into the observed region. Alternatively, they may have undergone binary interactions that significantly altered their apparent ages and evolutionary states. Even under this restricted scenario—assuming the RSGs and RGs represent the true evolved population and excluding any MS stars associated with the O stars—the observed number of MS stars remains far too low to match expectations. For example, the Poisson probabilities for Regions C, D, and E are $1.3 \times 10^{-3}$, $1.2 \times 10^{-2}$, and $7.8 \times 10^{-3}$, respectively. The combined probability of observing this pattern is just $1.2 \times 10^{-7}$, indicating that even this more conservative interpretation is strongly inconsistent with standard single-star evolution.

In summary, even if the O stars are unrelated to the surrounding stellar population -- as might be the case if they are runaways or binary products -- the observed MS population remains inconsistent with expectations from standard single-star evolution. Across a range of modeling assumptions, the combined distribution of bright evolved stars and MS stars cannot be reconciled with predictions from a coeval population under a standard IMF.     

As a more conservative alternative, we estimate the expected number of MS stars in Regions B through E (collectively, Region R) by conditioning on the observed number of evolved stars in Regions A, F, and G. Unlike the forward-likelihood approach, this method explicitly accounts for the uncertainty in small-number observations of the evolved populations. Although this yields a broader range of possible outcomes, it provides a more statistically robust test of consistency.

To illustrate the importance of this uncertainty, consider the two O stars in Region A. Based on the best-fitting model, we expect $\tilde{N}_{R|A} = 134.4$ MS stars in Region R. However, the Poisson uncertainty for two stars is large ($\sqrt{2} \approx 1.4$), , so it's entirely plausible that only one star was drawn from the underlying population. In that case, the expected MS count would drop to $\tilde{N}_{R | A} = 67.2$, which would be consistent with the observed value of $N_R = 72$. . This simple example highlights how small-number uncertainties in the evolved populations can dominate the inference. Therefore, to properly assess consistency, we estimate the posterior distribution of $\tilde{N}_R$given the observed numbers of O stars, RSGs, and RGs.

To formalize this approach, we estimate the posterior distribution for the total expected number of MS stars in Region R, $\tilde{N}_{R}$ conditioned on the observed numbers of O stars ($N_A$), RSGs ($N_{F}$), and RGs ($N_{G}$): 
\begin{equation}
P(\tilde{N}_{R}| N_A,N_F,N_G) = \frac{P(N_A,N_F,N_G|\tilde{N}_R)P(\tilde{N}_R)}{P(N_A,N_F,N_G)} \, .
\end{equation}
We model the likelihood as the product of three independent Poisson terms, each corresponding to an evolved stellar population whose size reflects a latent contribution to Region R:
\begin{multline}
P(N_A, N_F, N_G \mid \tilde{N}_R) = \\
P(N_A \mid \tilde{N}_A(\tilde{N}_{R|A})) \\
\times P(N_F \mid \tilde{N}_F(\tilde{N}_{R|F})) \\
\times P(N_G \mid \tilde{N}_G(\tilde{N}_{R|G})) \, .
\end{multline}
where $\tilde{N}_{R|A} + \tilde{N}_{R|F} + \tilde{N}_{R|G} = \tilde{N}_R $, and each term represents the expected contribution to Region R from one stellar age group. We define the expected number of O stars $\tilde{N}_A$ using a scaling relation from the Stellar Ages model:
\begin{equation}
\tilde{N}_A = \tilde{N}_{R|A}\frac{\tilde{N}_{A,S}}{\tilde{N}_{R|A,S}} \, .
\end{equation}
where $\tilde{N_{A,S}}$ and $\tilde{N_{R|A,S}}$ are the number of O stars and associated MS stars inferred by the Stellar Ages model. This ratio encapsulates the relationship between the observed evolved and MS stars in the model population. We calculate the likelihood $P(N_A|\tilde{N}_A)$ assuming a Poisson distribution centered at $\tilde{N}_A$, and similarly evaluate  $P(N_F | \tilde{N}_F)$ and $P(N_G| \tilde{N}_G)$ using the same approach.

\begin{figure}
\includegraphics[width=\columnwidth]{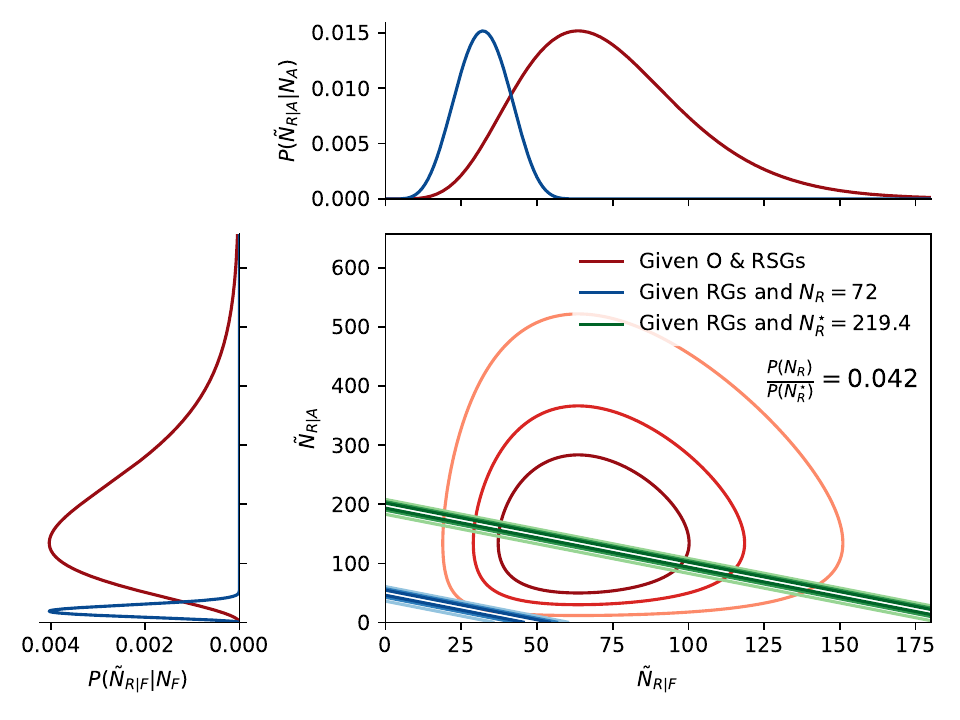}
\caption{Posterior distribution for the expected number of MS stars in Region R, defined as the union of Regions B, C, D, and E (see Figure~\ref{fig:threeages}). $\tilde{N}_{R|A}$, $\tilde{N}_{R|F}$, and $\tilde{N}_{R|G}$ represent the expected number of MS stars in Region R associated with the O stars (Region A), RSGs (Region F), and RGs (Region G), respectively. The bottom-right panel shows the posterior given the O and RSG stars (red contours) and the posterior given the RGs under the constraint that the total number of MS stars equals the observed value $N_R = 72$ (blue contours). Contours represent the 50\%, 75\%, and 95\% highest density credible intervals (HDCIs). The constraint $N_R = \tilde{N}_{R|A} + \tilde{N}_{R|F} + \tilde{N}_{R|G} = 72$ is in significant tension with model expectations. To quantify this tension, we compute the Bayesian evidence under this constraint and compare it to the model evidence under the most internally consistent case, where the expected number of MS stars is $N_R^{\star} = 219.4$. The resulting support ratio, $P(N_R)/P(N_R^{\star}) = 0.042$, indicates substantial tension between the model and the data. The top and left panels show marginalized posteriors: the red curves correspond to the O and RSG stars, while the blue curves represent the data-constrained posterior obtained by convolving $P(\tilde{N}_{R|A}, \tilde{N}_{R|F})$ with the RG term $P(\tilde{N}_{R|G} = N_R - \tilde{N}_{R|A} - \tilde{N}_{R|F})$.}
\label{fig:expectedposterior}
\end{figure}

The resulting posterior depends on three variables: \(\tilde{N}_{R|A}\), \(\tilde{N}_{R|F}\), and \(\tilde{N}_{R|G}\), which represent the expected number of MS stars in Region \(R\) associated with the O stars (Region A), RSGs (Region F), and RGs (Region G), respectively. These variables are constrained by the total number of observed stars in Region \(R\):
\begin{equation}
N_R = \tilde{N}_{R|A} + \tilde{N}_{R|F} + \tilde{N}_{R|G} \, .
\label{eq:constraint}
\end{equation}
Figure~\ref{fig:expectedposterior} presents the posterior distribution. Because the full posterior is three-dimensional, we use the constraint in eq.~(\ref{eq:constraint}) to visualize it via two overlapping two-dimensional projections. The red contours show the posterior conditioned on the stars in Regions A and F, while the blue contours represent the posterior conditioned on the stars in Region G. In the latter case, we use Eq.~\ref{eq:constraint} to express the third model variable, \(\tilde{N}_{R|G}\), in terms of the other two: \(\tilde{N}_{R|G} = N_R - \tilde{N}_{R|A} - \tilde{N}_{R|F}\). The product of these two distributions defines the full posterior constrained by \(N_R\). The contours correspond to the 50\%, 75\%, and 95\% highest-density credible intervals (HDCIs).

To quantify how well the model fits the observed MS population, we compute the Bayesian model evidence by integrating over the allowed region of the posterior under the constraint:
\begin{multline}
P(N_R) = \int \int P(\tilde{N}_{R|A}, \tilde{N}_{R|F}) \\
\times P\left(\tilde{N}_{R|G} = N_R - \tilde{N}_{R|A} - \tilde{N}_{R|F} \right) \, d\tilde{N}_{R|A} \, d\tilde{N}_{R|F} \, .
\end{multline}

We compare this to the model evidence in the most internally consistent case, where \(N_R^{\star} = 219.4\) is the number of MS stars predicted by the MIST models given the observed numbers of O stars, RSGs, and RGs. The corresponding constraint, 
\(P\left(\tilde{N}_{R|G} = N_R^{\star} - \tilde{N}_{R|A} - \tilde{N}_{R|F}\right)\), 
shown in green, overlaps the peak of the unconstrained posterior 
\(P(\tilde{N}_{R|A}, \tilde{N}_{R|F})\). However, the data-constrained posterior—defined by the observed number of MS stars—peaks largely outside this overlap region, lying mostly beyond the 95\% highest-density credible interval (HDCI) of the model-consistent posterior.

The ratio \(P(N_R)/P(N_R^{\star}) = 0.042\) quantifies the relative support for the observed MS star count versus the most model-consistent count. This is not a frequentist \(p\)-value, but a Bayesian measure of model adequacy. A support ratio this low indicates significant tension between the model and the data. Notably, despite several free parameters in the Stellar Ages model, the best agreement it can produce yields a support ratio below 0.05.

Given the low improbability of the missing MS stars under single-star evolution, another possibility is that they were originally present but have since migrated out of the region. If many of the missing stars were runaway stars, their kinematics should reflect this scenario.

\begin{figure}
\includegraphics[width=\columnwidth]{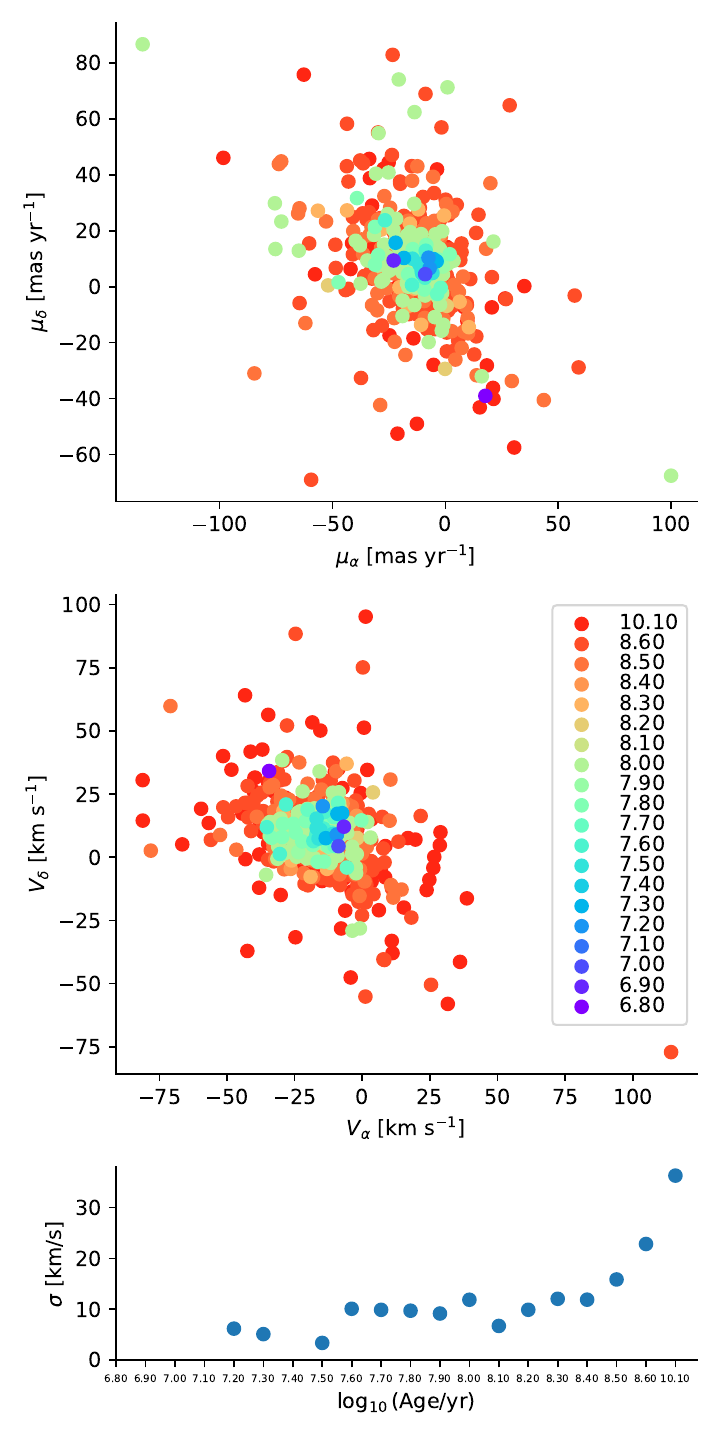}
\caption{
The transverse proper motions (top panel), transverse space velocities (middle panel), and velocity dispersion (bottom panel) for all stars with $M_{\rm RP}$ and $M_{\rm BP} < 0$ within 150 pc of the Vela Pulsar. The color scale represents the inferred ages of individual stars. As noted in previous studies, the proper motions and space velocities of the Vela-Puppis OB associations are indistinguishable from those of field stars. As expected from dynamical heating, older stars exhibit higher velocity dispersion. Crucially, there is no evidence for a significant population of runaway stars in this region.}
\label{fig:transverse}
\end{figure}

Figure~\ref{fig:transverse} shows the transverse proper motions (top panel), transverse space velocities (middle panel), and velocity dispersion (lower panel) for all stars within 150 pc of the Vela Pulsar with $M_{\rm BP}$ and $M_{\rm RP} < 0$. One way to distinguish stellar populations is by analyzing their distribution in phase space (both spatial and velocity space). However, as \citet{cantat-gaudin2019} noted, the proper motions of OB associations in the Vela-Puppis complex are indistinguishable from those of field stars. This is one of the key reasons we model the entire population collectively rather than attempting to isolate the Vela-Puppis OB associations. Given this significant kinematic overlap, any attempt to separate the OB stars from field stars would introduce selection biases that are difficult to model accurately.

The lower panel of Figure~\ref{fig:transverse} confirms a well-known dynamical trend: velocity dispersion increases with stellar age due to dynamical relaxation. More importantly, all of the young-looking stars exhibit relatively low space velocities compared to the surrounding field stars. This strongly suggests that the missing MS stars are not simply runaway stars that have migrated out of the region.

Taken together, these results rule out both standard single-star evolution and mass-dependent loss mechanisms, such as runaway stars, as viable explanations for the missing MS stars. Instead, the observed population strongly suggests that alternative evolutionary processes—such as binary interactions or extreme rotation—must have played a significant role. The following sections explore how these effects may have influenced the most luminous members of the Vela-Puppis complex.

\subsection{Velar Pulsar Progenitor Mass: Assuming that the RSGs are Siblings of the Progenitor}

A crucial question in understanding the Vela Pulsar’s origins is determining the mass of its progenitor star. The inferred age distribution of the surrounding stellar population suggests a mixed-age environment, making it challenging to identify which stars are most likely to have been coeval with the progenitor. However, {\it Stellar Ages} provides a unique advantage: it infers ages not just for the population as a whole, but also for individual stars. This allows us to use the brightest evolved stars as direct tracers of the progenitor, refining our inference beyond what is possible with population-level statistics alone. Given that red supergiants (RSGs) are the evolved descendants of massive stars, we estimate the progenitor mass by comparing it to the properties of the brightest RSGs in the region.

\begin{figure}
\centering
\includegraphics[width=\columnwidth]{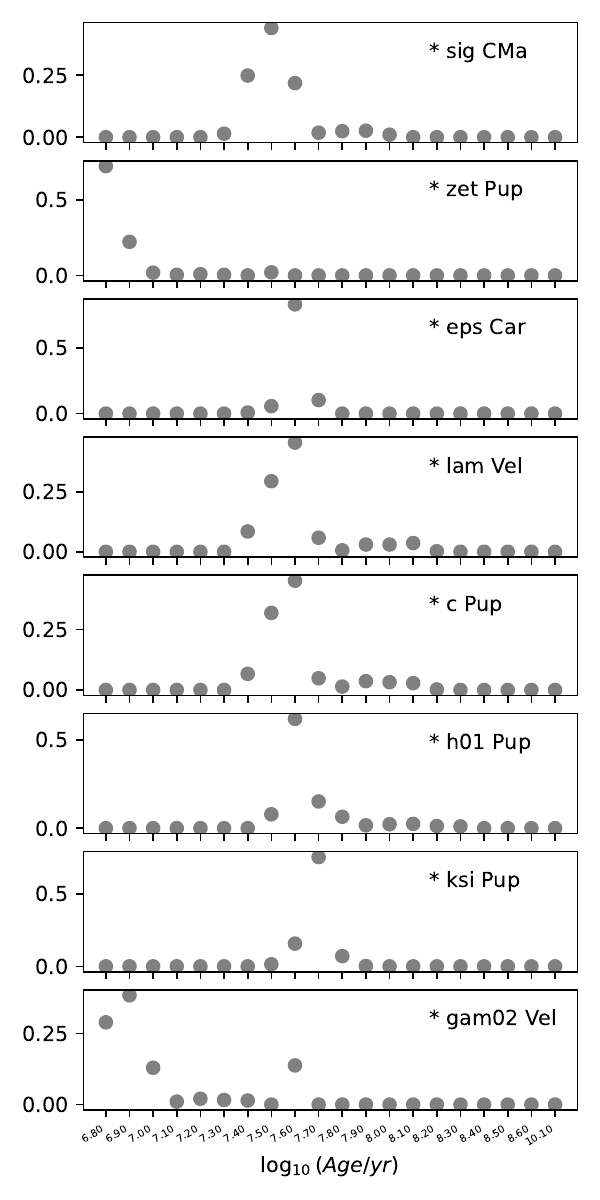}
\caption{Posterior age distribution for the eight brightest stars in the Gaia RP band. The six red supergiants (RSGs) share similar inferred ages, while the two peculiar O stars are clear outliers, suggesting that they did not follow the same evolutionary path as the RSGs. Their origins remain uncertain and could be explained by a different coeval population, multiple-star evolution, or very rapid rotation.
\label{fig:nbrightages}}
\end{figure}

\begin{figure}
\centering
\includegraphics[width=\columnwidth]{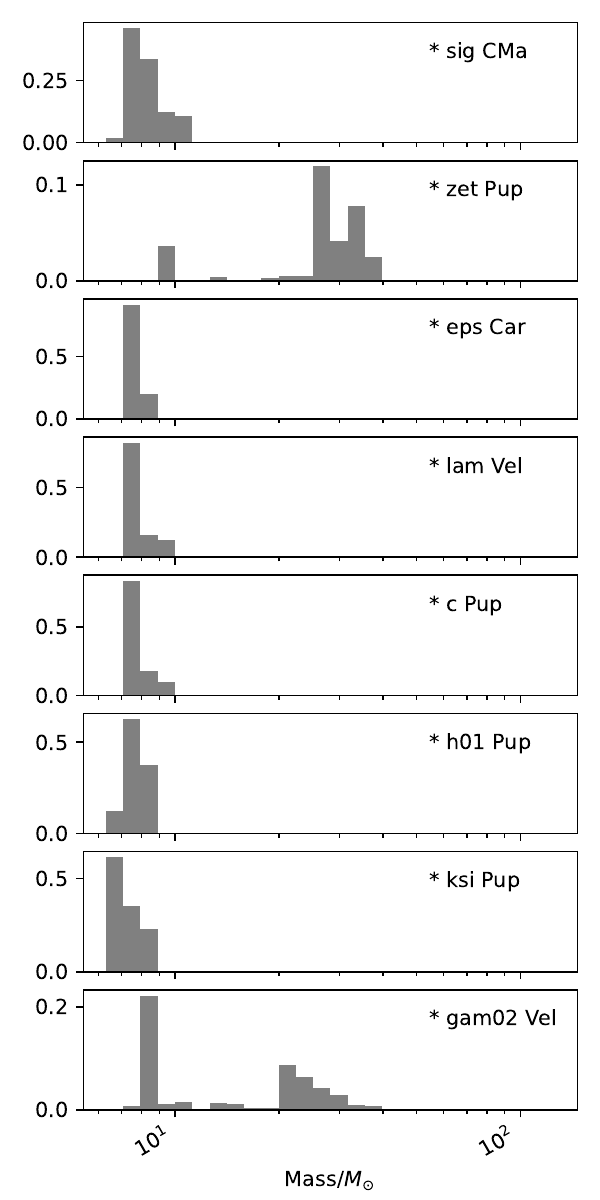}
\caption{Posterior mass distribution for the eight brightest stars in the Gaia RP band. The six red supergiants (RSGs) cluster around $\sim$8 M$_\odot$, while the two peculiar O stars have significantly higher inferred masses, further distinguishing them from the RSG population.}\label{fig:nbrightmasses}
\end{figure}

\begin{figure}
\centering
\includegraphics[width=\columnwidth]{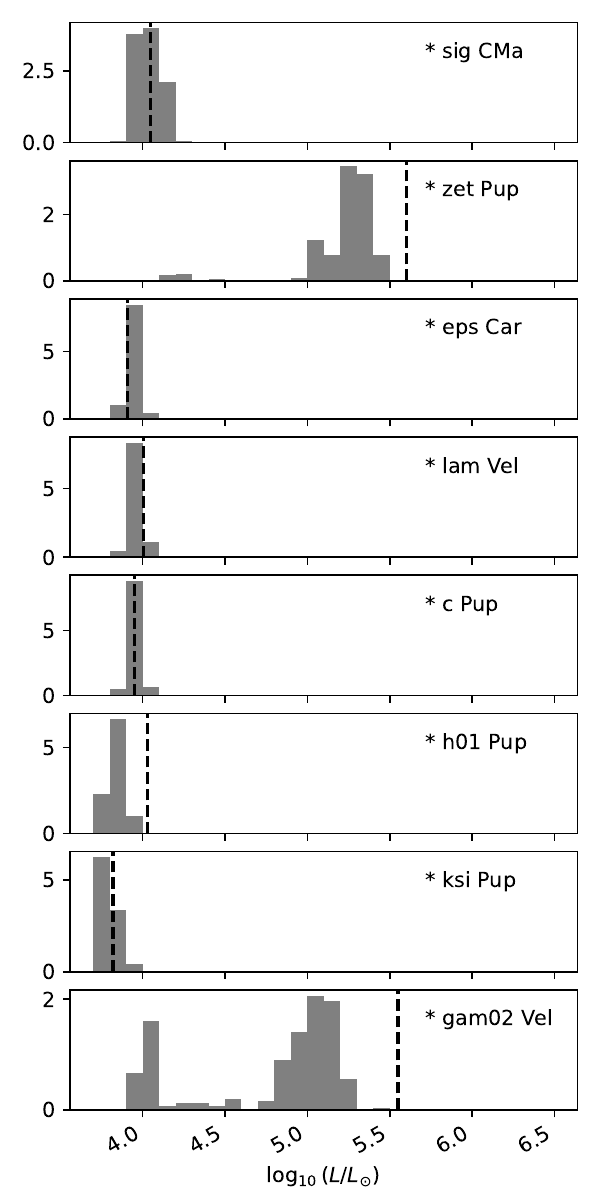}
\caption{
Posterior luminosity distribution for the eight brightest stars in the Gaia RP band. The vertical dashed lines indicate luminosity estimates based on spectral and luminosity classifications, allowing for comparison between the inferred and cataloged values.
}\label{fig:nbrightlogl}
\end{figure}

\begin{figure}
\centering
\includegraphics[width=\columnwidth]{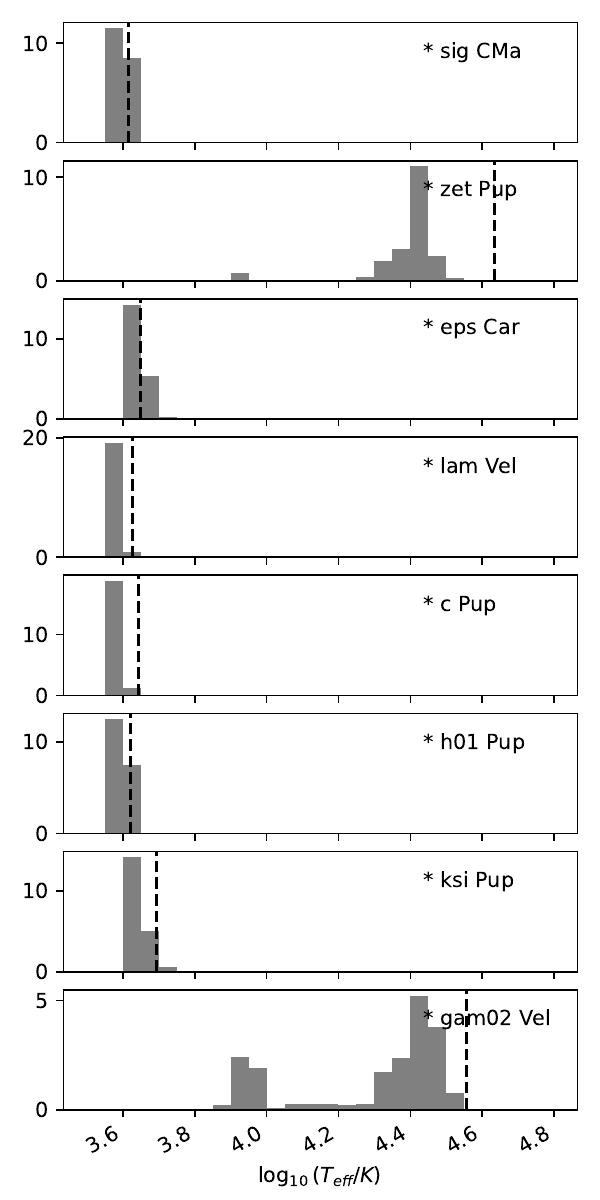}
\caption{
Posterior temperature distribution for the eight brightest stars in the Gaia RP band. The vertical dashed lines indicate temperature estimates based on spectral and luminosity classifications, providing a comparison between the inferred and cataloged values.
}\label{fig:nbrightlogt}
\end{figure}

Figures~\ref{fig:nbrightages}–\ref{fig:nbrightlogt} present the inferred ages, masses, luminosities, and effective temperatures for the eight brightest evolved stars in the sample. The vertical dashed lines in Figures~\ref{fig:nbrightlogl} and \ref{fig:nbrightlogt} indicate the luminosity and effective temperature estimates based on spectral and luminosity classifications from SIMBAD. The \textit{Stellar Ages} inferences are consistent with the SIMBAD properties for all six red supergiants (RSGs) and show partial agreement for the two peculiar O stars. Under standard single-star assumptions, the two O stars would have ages younger than 10 Myr, masses above 20 M$_\odot$, and correspondingly high luminosities and effective temperatures. In contrast, the six RSGs have inferred ages of $\log_{10}(t/{\rm yr}) = 7.6$ (~40 Myr) and masses around 8 M$_\odot$.  
However, as we show in Figures~\ref{fig:threeages}~\&~\ref{fig:expectedposterior}, there are not enough MS stars to support two distinct age populations under standard assumptions for stellar population modeling. While the differing inferred ages of the RSGs and O stars might suggest separate star formation episodes, this interpretation is inconsistent with the observed MS population. This tension implies that at least one standard assumption — such as single-star evolution, a constant IMF, or a uniform SFH — may not hold.

\begin{figure}
\centering
\includegraphics[width=\columnwidth]{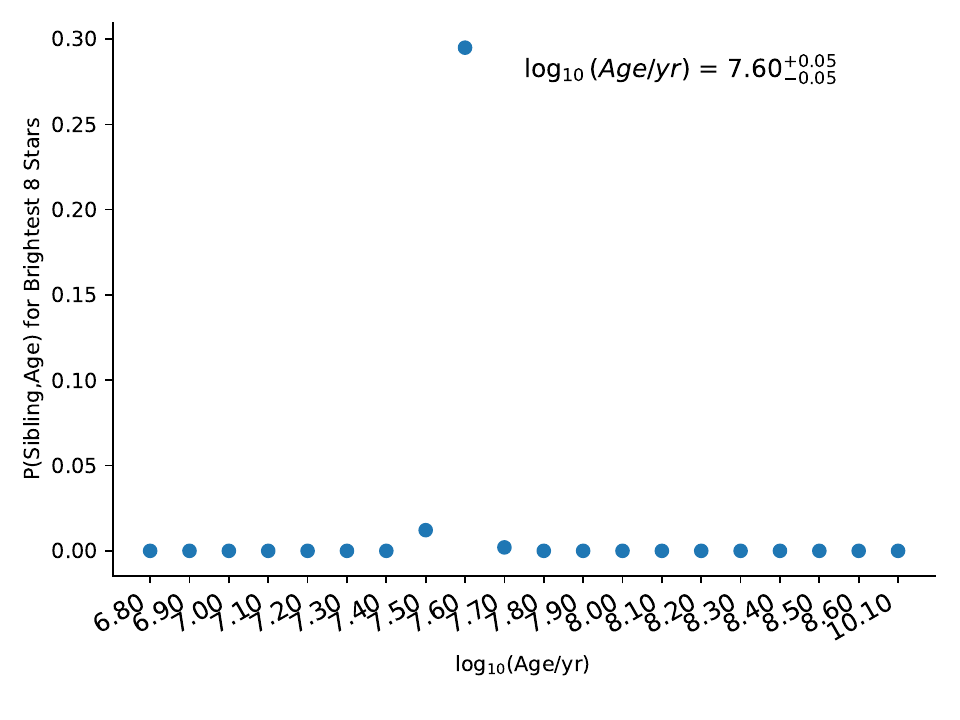}
\includegraphics[width=\columnwidth]{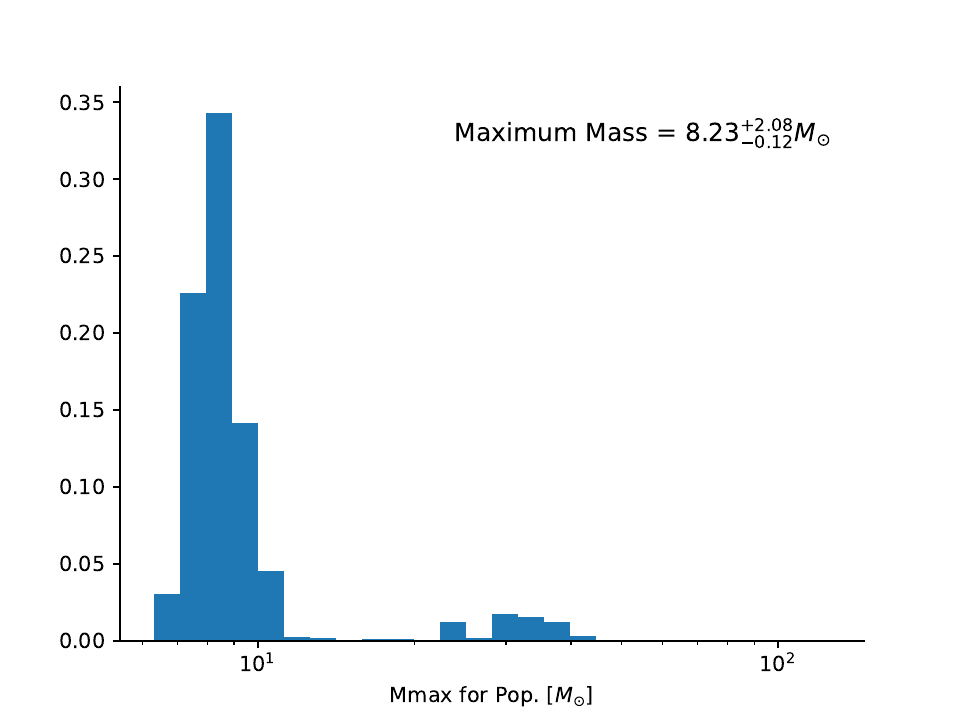}
\caption{
The top panel shows the posterior age distribution for the eight brightest stars in the Gaia RP band. This distribution is constructed by selecting MCMC draws in which at least half of these stars are coeval siblings (i.e., have the same inferred age). The bottom panel shows the corresponding maximum stellar mass for this age, assuming single-star evolution. The inferred age and mass distributions are dominated by the six RSGs, reflecting their strong influence on the coeval population assumption.
}\label{fig:siblings}
\end{figure}

Figure~\ref{fig:siblings} presents the inferred age and mass of the Vela Pulsar’s progenitor, under the assumption that at least half of the eight brightest stars are coeval siblings. The top panel shows the age distribution given that at least half of the eight brightest stars are coeval siblings. This is determined by selecting MCMC samples in which at least half of these stars share the same inferred age. Assuming standard single-star evolution, the inferred age of the brightest stars—and by extension, the progenitor -- is $\log_{10}(t/{\rm yr}) = 7.6 \pm 0.05$. This result is strongly driven by the six RSGs, which dominate the inference. The bottom panel shows the maximum stellar mass at this age, representing the upper mass limit for stars that evolved similarly to the RSGs. If the progenitor was indeed a sibling of these RSGs, then its initial mass—determining its evolutionary fate—was $M_{\text{max}} = 8.23^{+2.08}_{-0.12}$ M$_\odot$.

\subsection{Crossmatching the Young-looking Stars with the SIMBAD Database}
\label{sec:simbad}

To further investigate the nature of the young-looking ($\log_{10}(t/{\rm yr}) \le 7.7$) stars identified in Section~\ref{sec:results}, we cross-match their properties with classifications from SIMBAD. This cross-matching provides an independent way to assess whether their absolute magnitudes and spectral types are consistent with their inferred ages. Notably, some of these stars exhibit features suggesting they may have evolved through nonstandard pathways, such as multiple-star evolution or rapid rotation.

Table~\ref{tab:stellarproperties} lists the SIMBAD classifications for the young-looking stars identified in Section~\ref{sec:results}. Many of these stars are classified as Be stars, rapidly rotating B-type stars, or peculiar O stars. Some have previously been noted in the literature for their unusual properties, including high luminosities and inferred youth. The sample also contains a mix of luminous B-type stars and evolved blue stars, whose properties differ from standard main-sequence stars of similar absolute magnitudes.

\begin{table*}
\centering
\caption{Stellar properties of young-looking stars with inferred ages $\log_{10}(t/{\rm yr}) \leq 7.7$. The Name, Object Type (Object), and Spectral Type (Sp Type) are reported from the SIMBAD database after crossmatching, confirming the GSP-Phot inferences and supporting the extinction estimates. Among the most notable objects, $\zeta$ Puppis is an evolved O-type supergiant with strong winds and rapid rotation, $\gamma^2$ Vel (Regor) is a spectroscopic binary composed of a Wolf-Rayet star and a blue giant, and $\epsilon$ Car is a binary system consisting of a K giant and a B-type main-sequence star. Additionally, eight of the young-looking MS stars are Be stars, which are associated with rapid rotation, potentially due to binary evolution. Three stars are spectroscopic binaries (SB) or eclipsing binaries (EclBin), suggesting that a significant fraction of these stars may have undergone recent binary interactions. The columns $G_{BP}$ and $G_{RP}$ represent the Gaia absolute magnitudes, corrected for parallax, zero-point offsets, and extinction.}
\label{tab:stellarproperties}
\begin{tabular}{llllll}
\hline
\multicolumn{6}{|c|}{Gaia Young-looking Stars} \\
\hline
Name & Object & Sp Type & Gaia ID & BP & RP \\
\hline
* sig CMa &  RedSG &  K5Ib &  5610441600394646016 & -4.01 & -5.78\\
* c Pup &  SB* &  K2.5Ib-IIa &  5538814190283894656 & -3.74 & -5.53\\
* h01 Pup &  RedSG &  K4.5Ib &  5540591825707339648 & -3.49 & -5.25\\
* ksi Pup &  SB &  G6Iab-Ib &  5614471482318442240 & -3.96 & -5.23\\
* t02 Car &  Star &  K4/5(III) &  5350588691660896128 & -3.32 & -5.11\\
* omi Pup &  Be* &  B1IVe &  5602287587814648448 & -4.18 & -3.76\\
* kap CMa &  Be* &  B1.5Ve &  5583216215017531648 & -4.11 & -3.68\\
* r Pup &  Be* &  B1V &  5544672044638408832 & -4.00 & -3.63\\
* E Car &  Be* &  B3III &  5222191159719547776 & -3.70 & -3.45\\
V* V Pup &  SB* &  B1Vp+B2: &  5517171678276669696 & -3.47 & -3.25\\
* p Car &  Be* &  B4Vne &  5253796346624992128 & -3.53 & -3.21\\
* alf Pyx &  Variable* &  B1.5III &  5639188675501966464 & -3.44 & -3.18\\
* ome CMa &  Be* &  B2.5Ve &  5610257226031514240 & -3.36 & -3.18\\
* P Pup &  ** &  B0III &  5530670107652889344 & -3.55 & -3.18\\
* ups01 Pup &  Be* &  B2V+B3IVne &  5589411172768195840 & -3.42 & -3.10\\
HD  69081 &  Be* &  B2(V) &  5541637564345383808 & -3.19 & -2.92\\
* L Vel &  Star &  B2IV &  5312887640522116992 & -2.91 & -2.62\\
V* NO Vel &  SB* &  B3III &  5519315829032177152 & -2.77 & -2.59\\
HD  58286 &  PulsV* &  B2/3II/III &  5592847352767303680 & -2.69 & -2.46\\
* d03 Pup &  PulsV* &  B3III &  5586786020039670144 & -2.55 & -2.43\\
HD  66194 &  Be* &  B3Vn &  5290767631226220032 & -2.62 & -2.36\\
* b Pup &  SB* &  B2V &  5537733610870346880 & -2.56 & -2.32\\
V* IS Vel &  bCepV* &  B2V &  5516167755443050368 & -2.60 & -2.31\\
\hline
\multicolumn{6}{|c|}{Hipparcos Young-looking Stars} \\
\hline
Name & Object & Sp Type & HIP ID & BP & RP \\
\hline
* zet Pup &  BYDraV* &  O4I(n)fp &  39429 & -6.05 & -5.74\\
* eps Car &  ** &  K3:III+B2:V &  41037 & -4.34 & -5.62\\
* lam Vel &  RedSG &  K4Ib &  44816 & -3.73 & -5.54\\
* gam02 Vel &  WolfRayet* &  WC8+O7.5III-V &  39953 & -5.37 & -5.16\\
* kap Vel &  SB* &  B2IV &  45941 & -3.70 & -3.45\\
\hline
\end{tabular}
\end{table*}

Among the most notable cases in Table~\ref{tab:stellarproperties} are three peculiar bright, blue evolved stars. $\zeta$ Puppis is an evolved O-type supergiant with strong winds and rapid rotation. The other two, $\gamma^2$ Vel (Regor) and $\epsilon$ Car, are multiple star systems. $\gamma^2$ Vel is a spectroscopic binary consisting of a Wolf-Rayet star and a blue giant and is part of the quadruple star system $\gamma$ Vel. $\epsilon$ Car is a binary system containing a K giant and a B-type main-sequence star.

Additionally, nine of the young-looking stars are Be stars, a class known for rapid rotation and circumstellar disks.  Among the young-looking stellar systems, 20 contain at least one B star, and 9 of these host a Be star, yielding a Be fraction of 45\%. This is significantly higher than the average Be star frequency of 17\% \citep{zorec1997}, and somewhat above the typical value for B1/2 stars (30\%). Observational evidence, including strong emission lines, disk-like or oblate shapes from high-resolution interferometry, and Doppler broadening, suggests these stars rotate near breakup speeds \citep{porter2003}. Such extreme rotation may be a natural consequence of binary evolution \citep{pols1991,portegies1995,Shao2014,staritsin2022}. Alternatively, \citet{schneiderfabian2019} simulate a binary merger and find that the merger product can undergo internal angular momentum redistribution, resulting in slow surface rotation.  
In fact, nine of the stars in Table~\ref{tab:stellarproperties} are either visual doubles (**), confirmed spectroscopic binaries (SB), or eclipsing binaries (EclBin). In total, more than half of the young-looking stars—defined here as those with most likely ages $\log_{10}(t/{\rm yr}) \le 7.7$—exhibit signatures of either rapid rotation or binary evolution.

While these stars are undoubtedly members of the Vela-Puppis Complex, their age distribution has been a long-standing puzzle. Prior studies have noted inconsistencies in inferred stellar ages, raising the question of whether the Vela-Puppis population formed in a single event or across multiple star formation episodes. The following section summarizes these anomalies and places them in the broader context of the Vela-Puppis OB associations.

\subsection{The Mystery of the Vela-Puppis Complex of OB Associations}
\label{sec:mystery}

The stars analyzed in this work belong to the Vela-Puppis Complex, a vast network of OB associations spanning over 200 pc \citep{cantat-gaudin2018,cantat-gaudin2019}. While these stars are clearly part of this structure, their age distribution has been a long-standing mystery. Observational studies have repeatedly found discrepancies between expected and inferred stellar ages, suggesting a more complex star formation history than a single coeval burst. In an effort to resolve these inconsistencies, some studies have proposed that the region contains a mixture of multiple stellar populations with different ages. However, even allowing for a broad age range has not eliminated the anomalies, and significant inconsistencies remain. This section summarizes these studies and the known peculiarities of the Vela-Puppis population and its subgroups, placing them in their broader astrophysical context. The results of Section 4 further clarify these issues by robustly showing that the inconsistencies remain even when considering multiple ages.

The stars in this region span a wide range of sky positions and distances, encompassing several well-known OB associations that make up the Vela-Puppis Complex. These include Trumpler 10, Vela OB2, NGC 2547, BH 23, NGC 2451B, Collinder 135, UBC 7, and Collinder 140 (Cantat-Gaudin et al. 2018, 2019). Among these, the Vela Pulsar is closest to the Vela OB2 and Trumpler 10 associations, making them particularly relevant when considering possible connections between the pulsar’s progenitor and the surrounding stellar population.

Previous studies have used Pre-Main Sequence (Pre-MS) stars to estimate the ages of the OB associations in the Vela-Puppis Complex. By fitting Pre-MS isochrones to color-magnitude diagrams (CMDs), these studies have estimated that these associations span a broad age range from 10 Myr to 50 Myr \citep{cantat-gaudin2019}. While this suggests an extended period of star formation, it does not fully resolve the inconsistencies in inferred stellar ages.  Section~\ref{sec:results} shows that even when accounting for this range of ages, the observed brightness and mass distributions remain inconsistent with expectations.  

One particularly well-studied case is the region surrounding the bright O star $\gamma^2$ Vel. Using X-ray observations, \citet{pozzo2000} identified a group of Pre-MS, X-ray-active low-mass stars at the location of $\gamma^2$ Vel, which they termed the $\gamma$ Vela association. Subsequent optical studies by \citet{jeffries2009} detected a population of MS stars at the same location, but they noted a significant discrepancy in stellar counts: while the number of Pre-MS stars with masses between 0.1 and 0.6 M$_\odot$ is consistent with the expected number of stars in the 1.8–4.4 M$_\odot$ range, there are far too few stars to explain the presence of $\gamma^2$ Vel itself. This finding is consistent with our more formal statistical inference, which suggests that the brightest stars in this region cannot be easily explained under standard assumptions of single-star evolution.  

Age estimates for the $\gamma$ Vela group also show significant contradictions. While the Pre-MS CMD suggests an age of $\sim$10 Myr, later work by \citet{jeffries2017} found that lithium depletion measurements suggest a much older age of 20–30 Myr. This discrepancy further illustrates the ongoing difficulty in reconciling different age indicators, reinforcing the broader inconsistencies seen across the Vela-Puppis Complex.

The presence of two bright and peculiar O stars, $\zeta$ Puppis and $\gamma^2$ Vel, further complicates age inferences in the Vela-Puppis Complex. $\zeta$ Puppis is an evolved O-type supergiant (O4I(n)fp spectral classification) with strong winds and rapid rotation ($v_e \sin i = 213 \pm 7$ km s$^{-1}$). It exhibits photometric variations with a 1.78-day period, which could be due to stellar pulsation or rotation. If these variations are due to rotation, then $\zeta$ Puppis is rotating near break-up speed. Its brightness and effective temperature imply a mass of $\sim$50 M$_\odot$ if rotation is ignored, whereas models that include rotation prefer a lower mass of 20–30 M$_\odot$, with formal estimates ranging from 15 to 50 M$_\odot$ \citep{howarth2019}. As a result, the inferred age for $\zeta$ Puppis ranges from 5 to 15 Myr, depending on rotational effects.  

The other O star, $\gamma^2$ Vel, is a spectroscopic binary and part of the quadruple system Gamma Velorum. Using spectroscopy and astrometric solutions, \citet{north2007} provided the first complete orbital solution for $\gamma^2$ Vel, determining that the primary is an O-type giant (II-III luminosity class) with a dynamical mass of $28.5 \pm 1.1$ M$_\odot$, while the secondary is a Wolf-Rayet (WR) star with a dynamical mass of $9.0 \pm 0.6$ M$_\odot$. Evolutionary models suggest that the WR star originally had a mass of $\sim$30 M$_\odot$. \citet{eldridge2009} modeled the binary evolution of $\gamma^2$ Vel and inferred an age of $\sim$5 Myr. However, these models did not account for rotation, which, as demonstrated in the analysis of $\zeta$ Puppis, could significantly alter the inferred age.

In both cases, the evidence that these O stars have masses of approximately 20–30 M$_{\odot}$ is compelling. Under single-star evolution models, both stars suggest a young age. However, their mere presence within the Vela-Puppis complex remains perplexing. For decades, it has been noted that there are no other O stars within the complex, and more strikingly, their presence directly conflicts with the number of lower-mass stars and standard assumptions for the initial mass function (IMF).  

While this apparent discrepancy had been recognized before, \citet{jeffries2009} was the first to quantify this inconsistency for the $\gamma$ Velorum association specifically. The results of this study extend that analysis to a 150 pc radius region centered on the Vela Pulsar, demonstrating that the problem is not limited to a single association but rather a broader issue within the entire complex. Figure~\ref{fig:threeages} shows that assuming a Salpeter IMF, one would expect at least nine other MS O stars, yet none exist. Moreover, the observed number of MS stars is only half of what would be expected, a discrepancy that becomes even more severe when considering that these MS stars must also support the older populations at 40 Myr and 70 Myr.



\section{Possible Resolutions}
\label{sec:resolutions}

The observed stellar population within 150 pc of the Vela Pulsar is inconsistent with standard single-star evolution expectations. The number of main-sequence (MS) stars is too low to explain the presence of the observed red supergiants (RSGs) and bright evolved stars, assuming a standard initial mass function (IMF). Figure~\ref{fig:threeages} illustrates that, given a Salpeter IMF, at least nine additional MS O stars should exist in this region, yet none are found. Furthermore, the MS population is less than half of what is expected, and this discrepancy worsens when considering that the MS must also support both the 40 Myr and 70 Myr stellar populations. These inconsistencies demand an explanation.

Several possible resolutions could account for this discrepancy. First, the sample may be incomplete due to observational biases or missing stellar classifications. Second, mass-dependent dissolution of OB associations may have preferentially removed low-mass stars, artificially lowering the observed main-sequence population while leaving behind the more massive evolved stars. Third, the Vela Pulsar may be a run away star.  Fourth, the IMF itself may differ from the assumed Salpeter distribution, favoring more evolved stars at the expense of MS stars.  Fifth, a significant fraction of massive stars may be rotating at extreme velocities ($\omega / \omega_{\rm crit} \geq 0.9$), affecting their evolution and making them appear younger than they actually are. Finally, binary evolution could play a major role, as mass transfer, mergers, and binary-driven rejuvenation could significantly alter the expected number of MS stars and the evolved population.  The following evaluates each of these hypotheses in detail.

One possibility is that the stellar population within 150 pc of the Vela Pulsar is incomplete due to observational limitations. However, this is highly unlikely for the following reasons. The combination of Gaia and Hipparcos data should provide a nearly complete census of stars well below the magnitude limit. In particular, Gaia DR3 has high completeness for bright stars except for the very brightest, and supplementing with Hipparcos data ensures that the bright end of the population is fully accounted for. Our dataset therefore includes all stars within 150 pc down to absolute magnitudes of $M_{RP}$ and $M_{BP} = 0$. Given this depth, it is unlikely that missing stars could significantly alter the observed population.

Another possibility is that the dissolution of OB associations is mass-dependent, preferentially removing lower-mass MS stars while retaining the evolved, high-mass population. However, this scenario is unlikely given that our volume-limited sample covers a 150 pc radius, encompassing most, if not all, of the Vela-Puppis complex. If mass-dependent segregation were responsible, we would expect different effects for the low-mass (0.1–0.6 M${\odot}$) and intermediate-mass (1.8–4.4 M${\odot}$) populations. However, the ratio of Pre-MS to intermediate-mass MS stars is consistent with the Kroupa IMF. It seems implausible that mass segregation would dramatically affect the ratio of high-mass stars to lower-mass stars while leaving the ratio of intermediate-mass to low-mass stars unchanged.

Alternatively, one might consider that the Vela Pulsar progenitor was a runaway star, born in a different region and only later exploding locally. \citet{renzo2019} predict that massive stars can travel up to $\sim$100 pc before undergoing core collapse. However, given that our analysis includes all stars within a 150 pc radius, we are likely capturing the progenitor’s coeval stellar population. Additionally, this region encompasses most of the Vela-Puppis complex, and there are no other major OB associations within 100 pc that could plausibly be the progenitor’s birthplace.

Another argument against the runaway scenario is the pulsar’s relatively modest space velocity. While it is possible that the progenitor had an extremely high velocity, followed by a supernova kick in the opposite direction that nearly canceled its motion, such a scenario appears highly contrived. Given these considerations, we find it most likely that the Vela Pulsar’s progenitor originated within the stellar population analyzed in this study.

The initial mass function (IMF) is a fundamental assumption in modeling stellar populations. Could a variation in the IMF explain the observed discrepancies? \citet{kroupa2002} provided strong evidence that the IMF is largely universal, and since then, stellar population models have generally treated it as fixed. However, some studies have suggested that environmental conditions could lead to variations in the IMF \citep{bastian2010}. A meta-analysis by \citet{bastian2010} reviewed claims of nonuniversal IMFs and found that many could be explained by alternative interpretations rather than requiring a true IMF variation. They also noted that most IMF measurements lack robust statistical inference, making it difficult to consistently compare results across different studies.  

Even if an IMF variation were present, it is unlikely to resolve the observed inconsistencies. A top-heavy IMF, which produces an excess of massive stars, could, in principle, reduce the expected number of MS stars. However, this would not explain the simultaneous presence of evolved 4 M$_{\odot}$, 8 M$_{\odot}$, and 25 M$_{\odot}$ stars. Under standard assumptions, these evolved states have significantly different lifetimes, and a top-heavy IMF does not resolve this tension. Moreover, it would not naturally account for the apparent absence of MS O stars.  In contrast, binary evolution and rapid rotation provide a far more natural explanation for how 4 M$_{\odot}$ and 8 M$_{\odot}$ stars could appear in evolved stages, while also helping to explain the overall population inconsistencies. Nonetheless, even binary evolution and rapid rotation struggle to fully explain the presence of the two O stars, which remain the largest outstanding anomaly. Given these considerations, an IMF variation alone is highly unlikely to be the primary explanation for the observed discrepancies.

Another possible explanation is that a significant fraction of stars in this population are rotating at extremely high velocities. The rotation rates considered in this manuscript, however, are not rapid enough to fully explain the discrepancies between the data and the model. This limitation arises because the publicly available stellar models used here include only moderate rotation rates, with a typical value of $\omega / \omega_{\rm crit} = 0.4$, which represents the average rotation rate of massive stars in population synthesis models \citep{ekstroem2012}.

However, detailed studies of individual Be stars indicate a much broader range of rotation rates, with some stars reaching $\omega / \omega_{\rm crit} \sim 0.9$ \citep{porter2003}. Such extreme rotation could significantly alter the inferred stellar ages and luminosities for a given mass. Furthermore, most population synthesis models assume a single rotation rate for the entire population, but empirical studies of Be stars suggest that this does not accurately reflect the true distribution of stellar rotation rates.

Testing whether a broader range of rapid rotation rates could resolve the observed inconsistencies would require models that include both higher rotation rates and a wider distribution of rotational velocities than those currently available in publicly released stellar model grids.

Another possible resolution is binary evolution. Observations show that at least 90\% of B and O stars form in multiple-star systems (binaries, triples, etc.) \citep{offner2023}, and a large fraction of these will merge or exchange mass at some point during their evolution \citep{sana2012,demink2014}. Given this high fraction of interactions, it is natural to expect that binary evolution has played a major role in shaping the observed stellar population.  

One particularly compelling scenario is that the brightest RSGs and Be stars (regions F and C in Figure~\ref{fig:threeages}) are the products of binary evolution. Their inferred masses are roughly twice the mass of the brightest MS stars in a 65–100 Myr population, suggesting that they could have formed through mergers or mass transfer events. If this is the case, then the entire population could be explained self-consistently under a binary evolution framework. However, there are currently no publicly available isochrones that model this scenario, making direct comparisons challenging.  

The two peculiar O stars remain an open question. If these stars are truly very massive, then even binary evolution would struggle to explain their presence. However, it is possible that they are lower-mass stars with extremely rapid rotation. The origin of this rapid rotation remains a bit of a mystery.  \citet{schneiderfabian2019} showed that while the mergers of two stars leads to initial rotation near break-up, this very rapid rotation is short lived as the angular momentum transport and redistribution quickly slows the outer rotation down to $\sim$10\% of break-up.  There are some scenarios in which the merger of two compact, degenerate cores can lead to very rapid rotation, but such mergers are expected to be quite rare \citep{tout2011}.  

In summary, some combination of rapid rotation and binary evolution offers the most compelling resolution to the inconsistencies in the Vela-Puppis stellar population. Binary evolution naturally explains the discrepancy in the number of MS stars, the presence of evolved 4–8 M$_\odot$ stars, and the overabundance of Be stars and RSGs. While no binary evolution isochrones currently exist to formally test this scenario, future stellar population models incorporating mergers, mass transfer, and rapid rotation will be crucial for resolving these outstanding questions.

\section{Implications for CCSN Progenitor Estimates}
\label{sec:progenitor}

The study of core-collapse supernovae (CCSNe) progenitors is crucial for understanding which stars explode and why. There are three broad strategies to characterize CCSN progenitors. One approach is to model the supernova’s brightness evolution to infer the total ejecta mass or the size of the iron core that exploded \citep{bartunov1994,moriya2011,kasen2009,hillier2012,morozova2015,barker2023}. A second method is to analyze serendipitous pre-explosion imaging to model the brightness and color of the progenitor, allowing one to estimate the mass of the exploding star \citep{smartt2015,vandyk2017,strotjohann2023}. A third approach is to use stellar population age-dating to infer the zero-age main sequence (ZAMS) mass of the progenitor by modeling the brightnesses and colors of stars within $\sim$100 pc of the supernova’s location \citep{gogarten09a,williams2018,koplitz2023,diaz-rodriguez2021}.

Each of these methods provides unique insights but also relies on underlying assumptions about massive star evolution. The results of this paper directly impact all three approaches, highlighting the need to account for binary evolution and rotation when interpreting progenitor properties. More importantly, our analysis underscores that combining individual star inferences with stellar population analyses can yield significantly improved constraints on CCSN progenitors.

The age-dating technique has been widely used to infer the progenitor masses of CCSNe, particularly in cases where direct progenitor imaging or supernova observations are unavailable. When these constraints are missing, age-dating provides the only viable method for estimating the progenitor’s initial mass. For example, \citet{williams2018} inferred progenitor masses for all historical SNe within $\sim$8 Mpc, while \citet{jennings2014} used this approach to estimate the ages and masses of 115 SNRs. Based on these results, \citet{diaz-rodriguez2018} inferred a CCSN progenitor mass distribution, finding a minimum mass for CCSNe of $7.3 \pm 0.1$ M$_{\odot}$—currently the most precise statistical estimate of the minimum CCSN mass.

Age-dating is also one of the few methods available for inferring progenitor properties of compact remnants, including neutron stars. The Vela Pulsar is a prime candidate for this approach—it is relatively young, still embedded in its coeval stellar population, and has a well-constrained distance ($\sim$300 pc). \citet{kochanek2022} applied the age-dating technique to infer the most likely age and mass of the Vela Pulsar progenitor, using a standard Poisson-statistics-based CMD modeling approach. Like previous studies, they assumed single-star stellar evolution tracks to model the expected brightness and colors of the stellar population. Their analysis, incorporating Gaia magnitudes and parallaxes, along with nine bright Hipparcos stars missing from Gaia, found that the youngest age bin ($10^{7.4}$ – $10^{7.6}$ years) was most consistent with the CMD, implying a progenitor ZAMS mass of 8.1–10.3 M$_{\odot}$.

These previous analyses have relied on single-star models to interpret CMDs and infer progenitor properties. However, the results of this study suggest that this assumption may lead to systematic errors in the inferred age and evolutionary history of the progenitor. Because a star’s luminosity is closely tied to its mass, the inferred progenitor mass may still be reasonable. However, if a substantial fraction of massive stars undergo mergers or mass transfer, then the evolutionary pathways inferred from age-dating may not accurately reflect the progenitor’s true history. In particular, progenitors may appear to come from single-star evolutionary tracks when in reality they evolved through binary interactions or were influenced by rapid rotation. This suggests that while the inferred progenitor mass for the Vela Pulsar may still be reasonable, its evolutionary history and age estimates should be reconsidered in light of these findings.

A key takeaway from this study is that combining individual star analyses with population studies provides significantly more insight than either approach alone. Traditionally, progenitor properties have been inferred either by analyzing individual stars, such as through pre-explosion imaging, or by modeling the age distribution of a surrounding stellar population. While each method has its strengths, they also have significant limitations when used in isolation. Individual star analyses provide detailed constraints on a specific progenitor, but without context, they can miss the effects of binary evolution, rotation, and environmental influences. Conversely, population-based age-dating provides a broader statistical view, but assumptions about single-star evolution may lead to misleading results if the effects of binarity and rotation are not accounted for. By leveraging both approaches together—inferring properties of individual stars while also modeling the entire stellar population—we gain a more complete understanding of the evolutionary pathways leading to CCSNe. This study demonstrates how such an approach can identify discrepancies that might be overlooked by either method alone and reinforces the need for multi-faceted analyses in future progenitor studies.

The results of this study add to a growing body of evidence suggesting that binary evolution and rapid rotation are key factors in the evolution of massive stars leading to CCSNe. Prior observational and theoretical studies have demonstrated that a substantial fraction of CCSNe progenitors experience binary interaction at some stage of their evolution. For example, mass transfer in binary systems has been proposed as a major channel for stripped-envelope supernovae, including Type Ib and Ic SNe, where the hydrogen and/or helium envelope is removed before explosion. Additionally, binary evolution can produce mass gainers, mergers, or spun-up stars, which may lead to unexpected evolutionary pathways compared to standard single-star models. These findings align with the results of this study, which indicate that a significant fraction of the evolved massive stars in this population likely experienced binary interaction or rotational effects.

Spectroscopic observations show that at least 70\% of massive stars reside in binary systems \citep{sana2012}, and evolutionary modeling suggests that a similar fraction will have their evolution affected by mass transfer, with roughly 24\% eventually merging. These binary interactions can significantly alter the brightness and colors of evolved stars, directly affecting the stellar population surrounding a supernova. Recent work by \citet{menon2024} demonstrated that binary evolution substantially modifies the fraction of blue supergiants in a stellar population, reinforcing that binarity plays a critical role in shaping the observable characteristics of massive stars.

Several studies have specifically explored the impact of binarity on CCSN progenitors and their host stellar populations. \citet{zapartas2017} predicted that 15\% of CCSNe should be associated with relatively old stellar populations, while \citet{zapartas2019} suggested that 30–50\% of all Type II SNe progenitors likely underwent binary evolution. A key observational signature of binary evolution in CCSNe is a discrepancy between the inferred mass of the progenitor and the age of its surrounding stellar population \citep{zapartas2021, bostroem2023}. Single-star evolution predicts that the maximum age of the surrounding population should be at most $\sim$45 Myr, corresponding to the lifetime of the most massive stars that explode as Type II SNe. However, in a binary scenario, the merger of two $\sim$4 M$_{\odot}$ stars near the end of their lives ($\sim$100 Myr) could result in a rejuvenated, more massive progenitor that later undergoes core collapse. This provides a natural explanation for why some CCSNe should be associated with older stellar populations—yet a definitive observational confirmation of this effect remains elusive.

Similar age discrepancies have been observed in other young stellar populations containing red supergiants (RSGs). \citet{beasor2019} found that in four local stellar populations, the inferred ages from the brightest MS stars were around 10 Myr, whereas the inferred ages of the dimmest RSGs were closer to 20 Myr. This discrepancy was interpreted as evidence of either binary evolution or rapid rotation, further supporting the idea that single-star evolutionary models do not fully capture the diversity of massive star evolutionary pathways.

There is a growing body of direct and indirect evidence suggesting that binary evolution and/or rapid rotation significantly impact CCSN progenitors. One likely evolutionary pathway for Type Ib/c SNe is binary evolution, where mass transfer or stripping removes the progenitor's outer hydrogen envelope. Recent observations of peculiar light curves and spectral features in some Type Ib/c SNe further hint at interactions between the SN blast and pre-SN mass ejections, reinforcing the idea that binary evolution significantly shapes these explosions \citep{kuncarayakti2023, moore2023, agudo2023}.

Additional clues come from the search for surviving binary companions in SN remnants. \citet{fraser2019} used Gaia Data Release 2 to search for a possible companion to the Vela Pulsar, identifying a faint star kinematically consistent with being a runaway companion. However, they noted that this could also be an interloper, leaving the question open. More definitively, \citet{folatelli2016} and \citet{eldridge2016} studied the disappearance of a helium giant progenitor to the Type Ib supernova iPTF13bvn, concluding that the progenitor was most likely shaped by binary evolution.

The broader diversity of SN light curves also supports the role of binary evolution. \citet{eldridge2018} demonstrated that the variety of SN light curves naturally arises from binary evolution models, suggesting that binarity may be a key factor in shaping supernovae diversity. Similarly, \citet{xiao2019} found that HII regions ionized by binary-star models produce metallicities that better match independent measurements, further indicating that binarity alters the expected chemical evolution of star-forming environments. They also found that Type II and Type Ib/c SNe originate from similar age distributions, aligning with expectations from binary evolution models.

Together, these studies reinforce the conclusion that binary evolution and rapid rotation are not just minor corrections to massive star evolution, but fundamental processes that must be accounted for when interpreting CCSN progenitors.

\section{Conclusions}
\label{sec:conclusions}

This study reexamines the stellar population within 150 pc of the Vela Pulsar and finds that the standard assumptions in modeling stellar populations—particularly the assumption of single-star evolution—are inconsistent with the observations. Using Gaia parallaxes and the {\it Stellar Ages} algorithm, we infer a most likely progenitor mass of $8.2^{+2.1}_{-0.1}$ M$_{\odot}$, assuming that the Vela Pulsar’s progenitor was coeval with the 6 RSGs. However, the surrounding stellar population exhibits several anomalies that challenge traditional population synthesis methods. Notably, while there is very weak evidence for a population younger than 10 Myr and marginally weak evidence for a 40 Myr old population, the most statistically significant population is 65–100 Myr old. The inconsistency is that these ages are driven by the presence of three groups of evolved stars—O supergiants, red supergiants, and red giants—yet the corresponding main-sequence (MS) stars needed to support them are missing. This discrepancy not only complicates our understanding of the Vela Pulsar’s progenitor but also suggests fundamental issues in modeling the evolutionary history of the entire Vela-Puppis complex.

This severe mismatch between evolved stars and the expected MS population is one of the most striking inconsistencies in the stellar population. For example, the presence of the two peculiar O stars, $\zeta$ Puppis and $\gamma^2$ Vel, is entirely confounding. Assuming standard single-star evolution with modest rotation, one would expect at least nine other O stars within 150 pc, yet none are observed anywhere in the entire Vela-Puppis complex of OB associations. Furthermore, if these O stars formed through normal single-star evolution, their presence would imply a significantly larger population of main-sequence stars at lower masses, which is not observed. The red supergiants (RSGs) also present a major inconsistency, as their number does not match the expected number or luminosity distribution of main-sequence stars, further challenging standard evolutionary predictions. Taken together, the two peculiar O stars, the six RSGs, and the bright red giants suggest that there should be three times as many main-sequence stars as currently exist in this region.

This significant mismatch indicates that standard single-star evolution models alone cannot explain the observed population. Instead, binary evolution and/or very rapid rotation provide a more natural explanation. Binary interactions—through mass transfer, mergers, and rejuvenation—can alter evolutionary timescales, produce overluminous stars, and reduce the expected number of main-sequence stars. Similarly, rapid rotation—whether primordial, merger-induced, or the result of binary interactions—can modify stellar lifetimes and affect the observed distribution of stellar ages. Together, these effects offer a self-consistent alternative to standard single-star evolution models.

A key strength of this study is that it leverages the {\it Stellar Ages} method, which uniquely combines individual star inferences with population-level constraints. This combined approach enables a more nuanced analysis than traditional techniques, revealing internal inconsistencies with standard assumptions in modeling stellar populations. In particular, the results of {\it Stellar Ages} suggest that future studies should aim to incorporate more comprehensive binary evolution models and a wider range of rotational effects into stellar population synthesis methods. Expanding the available isochrone grids to include binary evolution tracks and extreme rotation cases will be essential for improving the accuracy of progenitor mass estimates. Ultimately, a more complete understanding of massive star evolution—one that fully accounts for the complexities revealed in this study—will be critical for making robust predictions about which stars explode, which collapse directly to black holes, and how stellar populations evolve over time.

\section*{Acknowledgments}
JWM was supported by the Los Alamos National Laboratory (LANL) through its Center for Space and Earth Science (CSES). CSES is funded by LANL’s Laboratory Directed Research and Development (LDRD) program under project number 20210528CR.  This work was supported by NASA through grant HST-GO-16778.017-A from the Space Telescope Science Institute, which is operated by the Association of Universities for Research in Astronomy, Inc., under NASA contract NAS 5-26555.  This work has made use of data from the European Space Agency (ESA) mission
{\it Gaia} (\url{https://www.cosmos.esa.int/gaia}), processed by the {\it Gaia}
Data Processing and Analysis Consortium (DPAC,
\url{https://www.cosmos.esa.int/web/gaia/dpac/consortium}). Funding for the DPAC
has been provided by national institutions, in particular the institutions
participating in the {\it Gaia} Multilateral Agreement.

%






\appendix

\section{Comparing Likelihoods from Different Models}

To test robustness of conclusions of this manuscript to differences in models, we infer the ages using two different models: MIST and Parsec1.2.  The primary part of this manuscript shows the results from the MIST models and this appendix first shows a comparison between the likelihoods and then shows the results from the Parsec1.2 models.

\begin{figure}
\begin{center}
\includegraphics[width=0.5\textwidth]{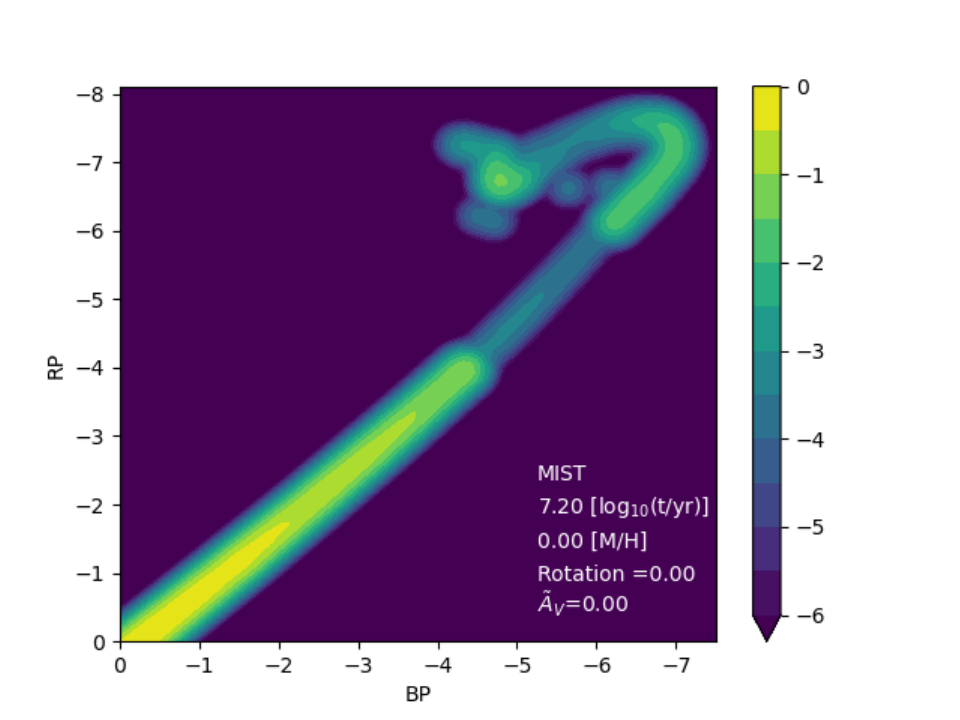}
\caption{Figures~\ref{fig:comparelikeli1}-\ref{fig:comparelikeli3} compare the likelihoods for different ages ($\log_{10}(t/yr) =$ 7.2 and 7.3), metaliticities ([M\H] = 0.0 and 0.2), models (MIST vs Parsec1.2), and rotation rates ($v_{\rm ini} = $ 0 and 0.4).  This Figure shows the likelihood for the MIST model with $\log_{10}(t/{\rm yr})=7.2$, [M/H] = 0.0, and $v_{\rm ini} = 0.0$; this likelihood is the reference model for all subsequent comparisons.}
\label{fig:comparelikeli1}
\end{center}
\end{figure}

\begin{figure}
\centering
\includegraphics[width=0.45\textwidth]{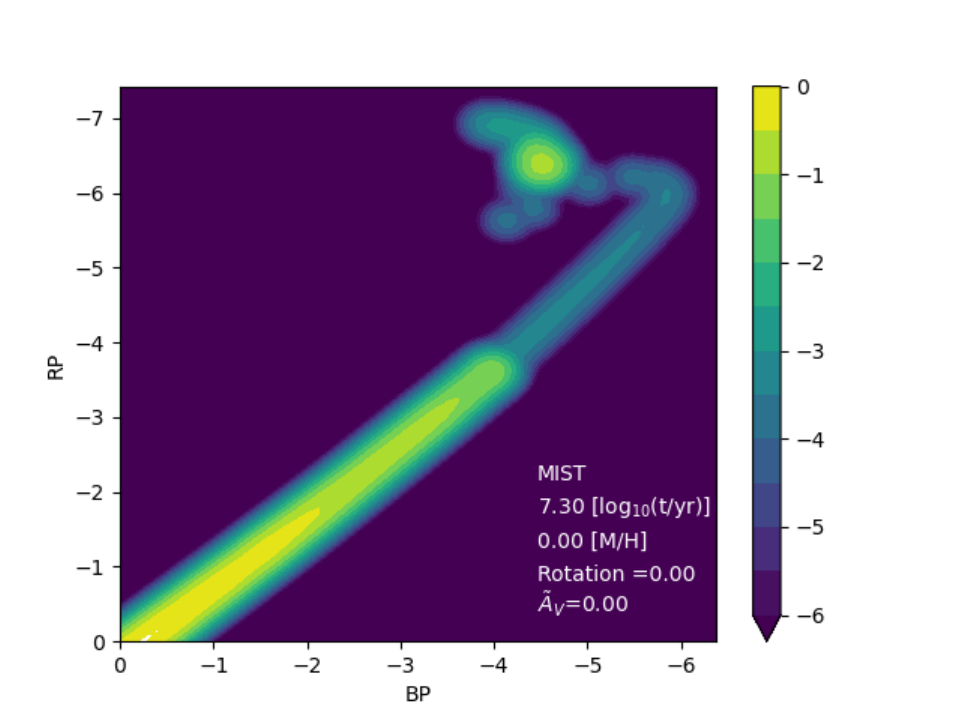}
\includegraphics[width=0.45\textwidth]{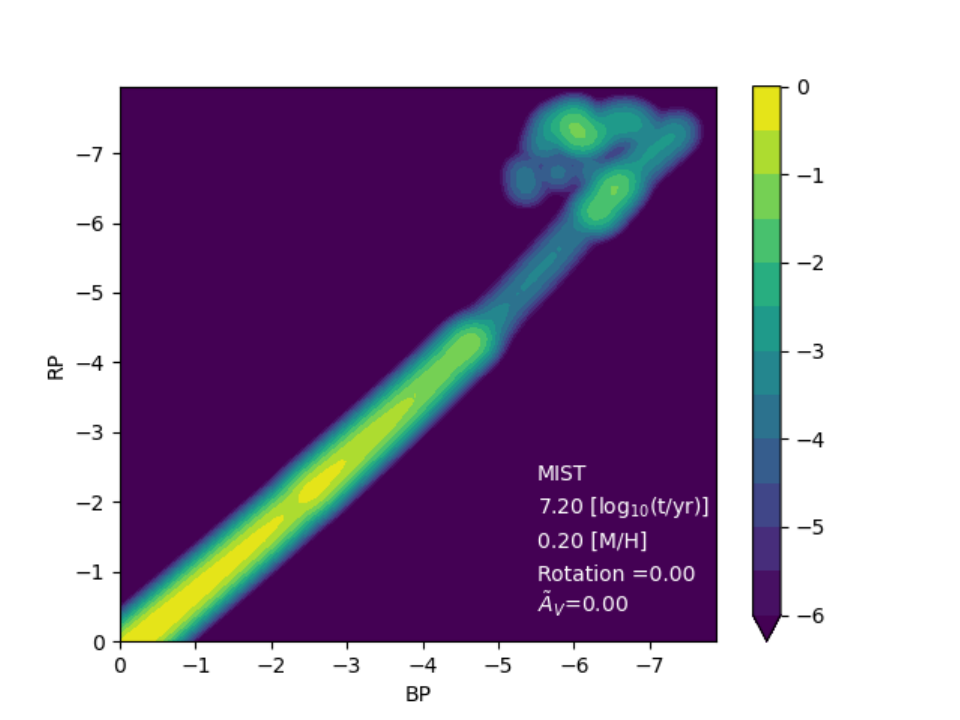}
\\
\includegraphics[width=0.45\textwidth]{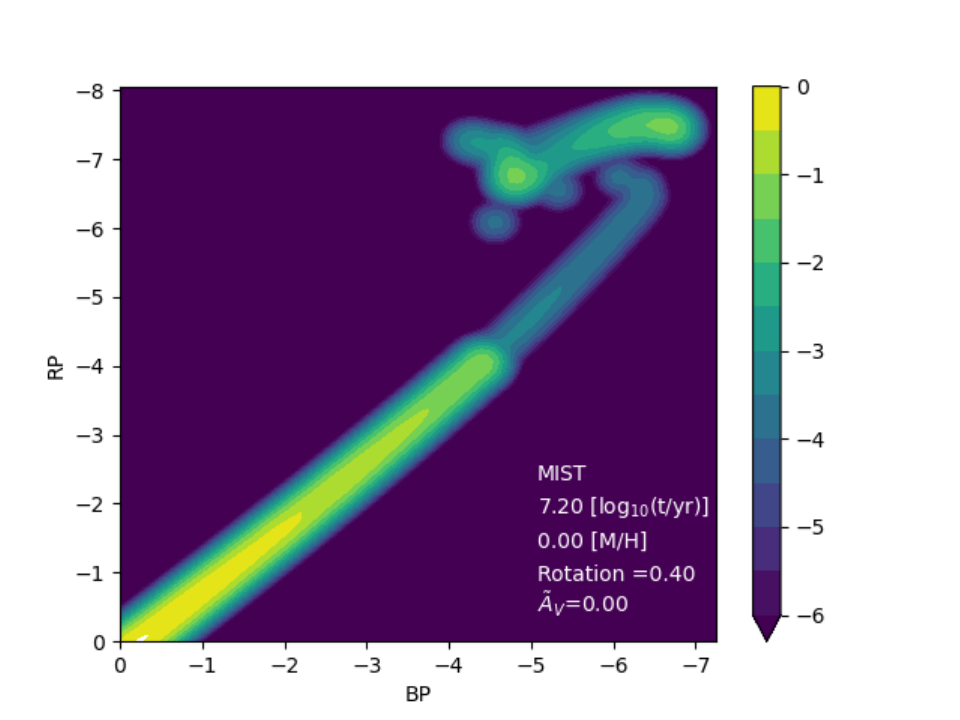}
\includegraphics[width=0.45\textwidth]{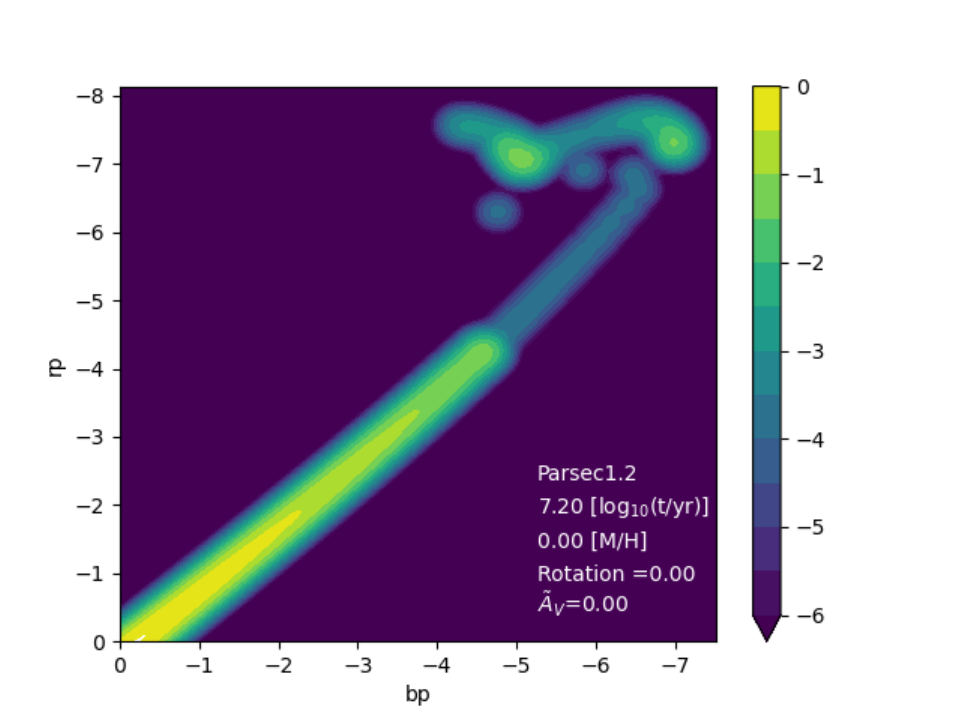}
\caption{Comparing the likelihoods for different parameters: age $\log_{10}(t/yr) = 7.3$ (a), metallicity [M/H] = 0.2 (b), rotation $v_{\rm ini} = 0.4$ (c), and model Parsec1.2 (d).}
\label{fig:comparelikeli2}
\end{figure}

\begin{figure}
\centering
\includegraphics[width=0.45\textwidth]{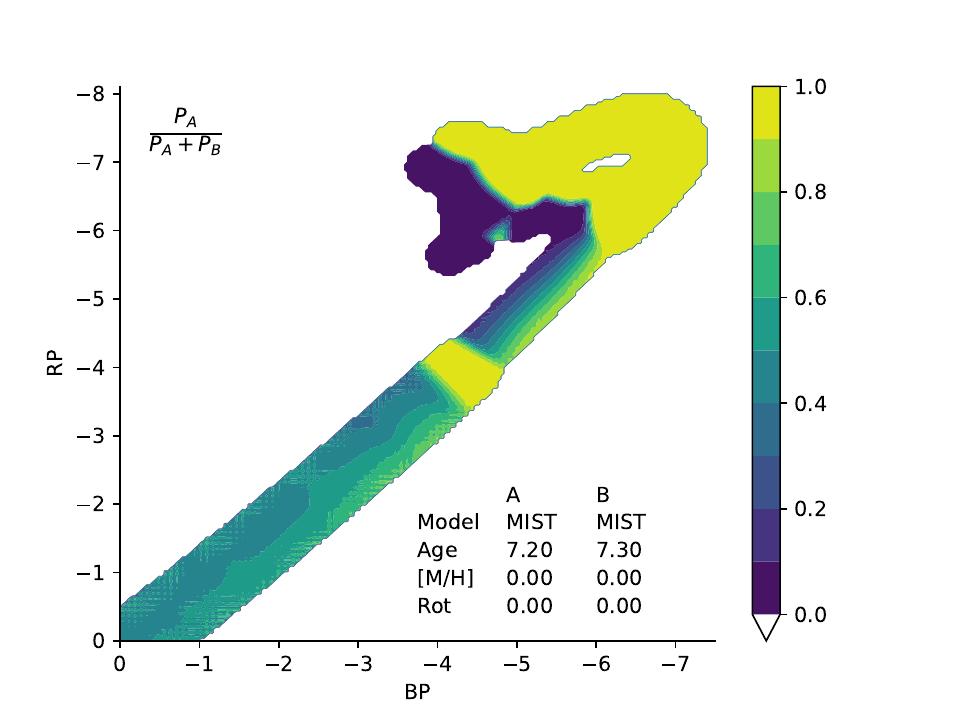}
\includegraphics[width=0.45\textwidth]{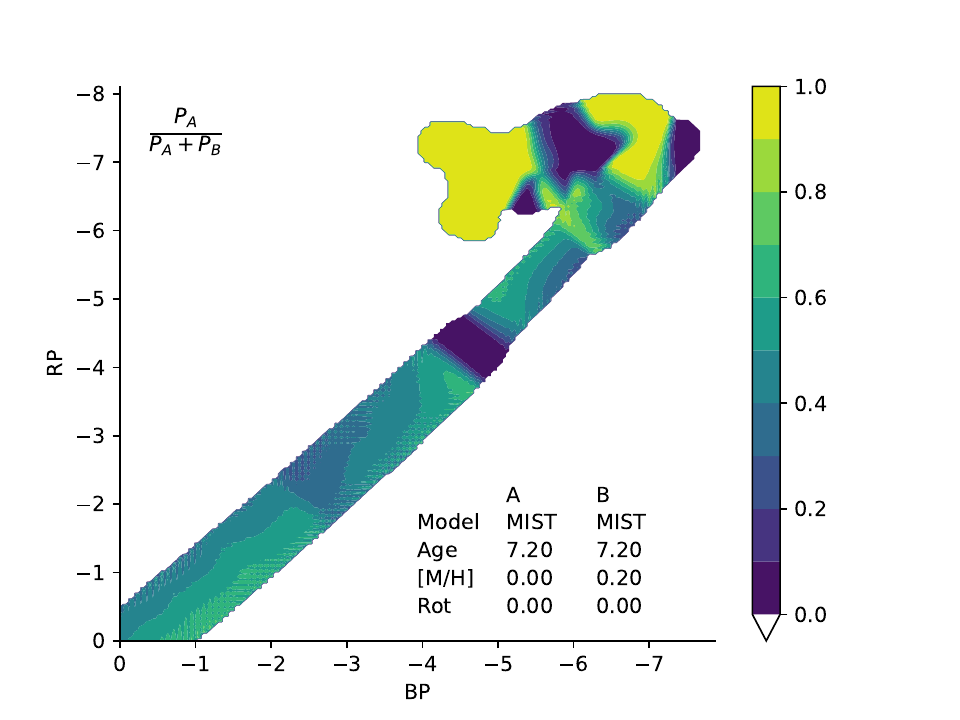}
\\
\includegraphics[width=0.45\textwidth]{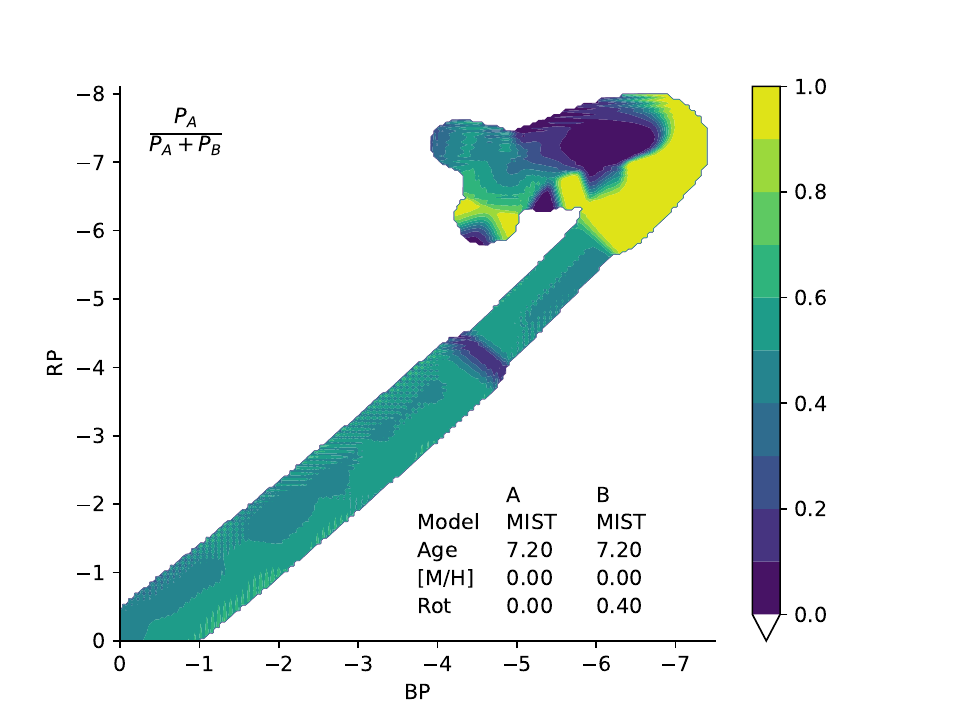}
\includegraphics[width=0.45\textwidth]{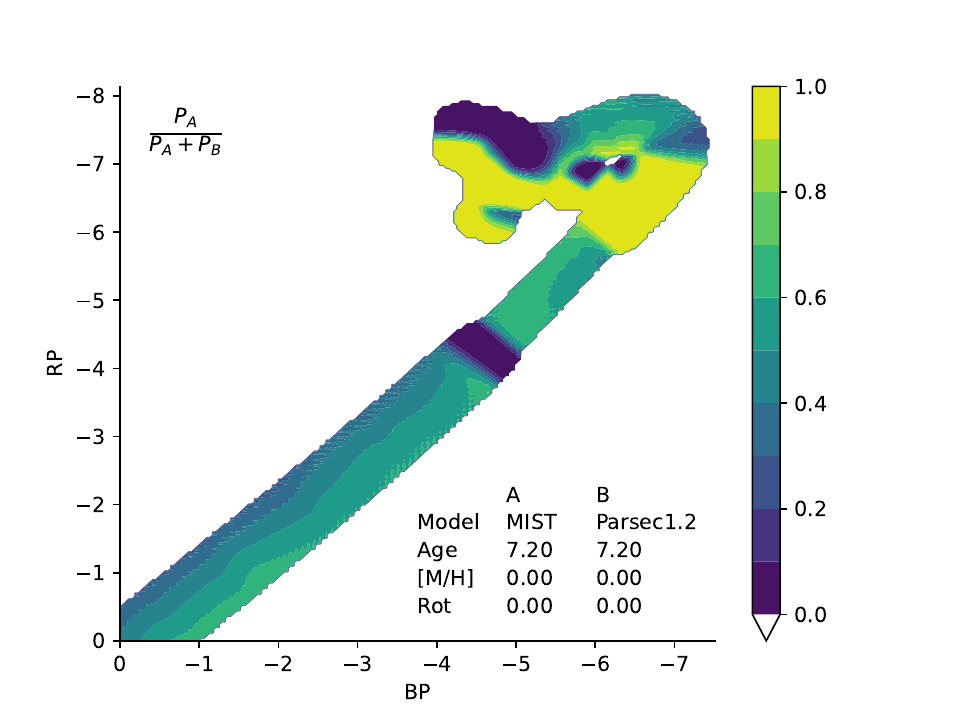}
\caption{A normalized comparison of the reference likelihood in Figure~\ref{fig:comparelikeli1} with each likelihood in Figure~\ref{fig:comparelikeli2}.  Each panel shows $P_A/(P_A + P_B)$, where $P_A$ is the reference likelihood.  Where the likelihoods are similar this ratio is 0.5, and if the reference likelihood dominates, then this ratio is 1.  Panel (a) compares ages ($\log_{10}(t/ {\rm yr})=7.2$ vs.~7.3); Panel (b) compares metallicities ([M/H]=0.0 vs.~0.2); Panel (c) compares initial rotation rates $v_{\rm ini} = 0.0$ vs.~0.4; Panel (d) compares different models (MIST vs.~Parsec1.2).  These comparisons show that a modest difference in age (0.1 dex) corresponds to larger differences in the likelihood than when comparing 0.2 dex in metallicity (panel b), rotation, or different models.  These comparisons also show that the differences in these four parameters manifest differently in the likelihoods.  Hence these four parameters are not generate and potentially distinguishable.}
\label{fig:comparelikeli3}
\end{figure}

Figures~\ref{fig:comparelikeli1}-\ref{fig:comparelikeli3} show comparisons in the likelihoods between MIST and Parsec1.2 models.   Figure~\ref{fig:comparelikeli1} is the reference model for these comparisons.  It shows the likelihood for a MIST model with $\log_{10}(t/{\rm yr}) = 7.2$, solar metallicity [M/H] = 0.0, and an initial rotation of $v_{\rm ini} = 0.0$.  {\it Stellar Ages} also includes a mode to infer the extinction distribution.  Since Vela Pulsar stars are relatively close and we have extinction estimates for all stars, there is no need to infer the extinction distribution.  As a result, the mean extinction $\tilde{A}_V = 0.0$ for this and all subsequent models.  Figure~\ref{fig:comparelikeli2} demonstrates the variation in likelihoods with a variation in age , metallicity, rotation, and model: panel (a) shows the likelihood for $\log_{10}(t/{\rm yr})=7.3$.  This age is 25\% older than the age ($\log_{10}(t/{\rm yr})=7.2$) in the reference model.  Panel (b) shows the likelihood for a metallicity at is 60\% higher than solar ([M/H] = 0.2).  Panel (c) shows a likelihood in which the initial rotation of the massive stars is 40\% of critical rotation ($v_{\rm} = 0.4$).  Panel (d) shows the likelihood for a Parsec1.2 model.

A useful way to compare two likelihoods ($P_A$ and $P_B$) is to show a normalized probability of the two Likelihoods: $P_A/(P_A+P_B)$.  Figure~\ref{fig:comparelikeli3} uses this normalized ratio to compare the likelihoods in Figure~\ref{fig:comparelikeli2} to the reference likelihood in Figure~\ref{fig:comparelikeli1}.  In all panels, the reference model is $P_A$.  When the likelihoods are similar, this normalized ratio is 0.5, and where the reference model dominates, the ratio is 1.0.  Conversely, this ratio is 0 where the comparison model, $P_B$, dominates.  The most striking result of these comparisons is that a modest difference in age (25\%) leads to the largest differences in the likelihood than using a different model, rotation, or metallicity.  Among the other parameters (metallicity, rotation, and model) the magnitude of differences is similar.  Intriguingly, the character of these differences is unique among these parameters.  Thus, an inference algorithm such as {\it Stellar Ages} can in principle distinguish among all four parameters.  Whether it can in practice depends upon observational uncertainties and whether these difference actually represent trends in Nature.

\bibliographystyle{aasjournalv}



\end{document}